%% file: 15TCOM_qmmimo_final.tex
\documentclass[journal,10pt,twoside,twocolumn]{IEEEtran}

\usepackage{cite}
\usepackage[T1]{fontenc}
\usepackage{graphicx}
\usepackage{amssymb}
\usepackage{amsmath}
\usepackage{paralist}
\usepackage{subfigure}
\usepackage{setspace}
\usepackage{microtype}
\usepackage{hyperref}
\usepackage{url}
\usepackage{balance}

\usepackage{algorithm}
\usepackage{algpseudocode}

\input{vmr-symbols-vecbold}
\input{standard-macros}

\input{defs}

\IEEEoverridecommandlockouts 

\usepackage{color}
\newcommand{\revision}[1]{#1}

\begin{document}

\title{Quantized Massive MU-MIMO-OFDM Uplink}

\author{ 
Christoph Studer,~\IEEEmembership{Senior Member,~IEEE}, and Giuseppe Durisi,~\IEEEmembership{Senior Member,~IEEE}
\thanks{C.~Studer is with the School of Electrical and Computer Engineering, Cornell University, Ithaca, NY (e-mail: studer@cornell.edu).}
\thanks{G.~Durisi is with the Department of Signals and Systems, Chalmers University, G\"oteborg, Sweden (e-mail: durisi@chalmers.se).}
\thanks{The work of C.~Studer was supported in part by Xilinx Inc.\ and by the US National Science Foundation (NSF) under grants ECCS-1408006 and CCF-1535897.}
\thanks{The work of G.~Durisi was supported by the Swedish Foundation for Strategic Research under grants SM13-0028 and ID14-0022, and by the Swedish Governmental Agency for Innovation Systems (VINNOVA) within the VINN Excellence center Chase.}
\thanks{\revision{The system simulator for the quantized massive MU-MIMO-OFDM uplink studied in this paper will be  available on GitHub: \url{https://github.com/quantizedmassivemimo/mu_mimo_ofdm/}}}
}

\maketitle

\begin{abstract}
Coarse quantization at the base station (BS) of a massive multi-user (MU) multiple-input multiple-output (MIMO) wireless system promises  significant power and cost savings.
Coarse quantization also enables significant reductions of the raw analog-to-digital converter (ADC) data that must be transferred from a spatially-separated antenna array to the baseband processing unit. 
The theoretical limits as well as practical transceiver algorithms for such quantized MU-MIMO systems operating over frequency-flat, narrowband channels have been studied extensively.
However, the practically relevant scenario where such communication systems operate over frequency-selective, wideband channels is less well understood. 
This paper investigates the uplink performance of a quantized massive MU-MIMO system that deploys orthogonal frequency-division multiplexing (OFDM) for wideband communication. 
We propose new algorithms for quantized maximum a-posteriori (MAP) channel estimation and data detection, and we study the associated performance/quantization trade-offs. 
Our results demonstrate that coarse quantization (e.g., four to six bits, depending on the ratio between the number of BS antennas and the number of users) in massive MU-MIMO-OFDM systems entails virtually no performance loss compared to the infinite-precision case at no additional cost in terms of baseband processing complexity.
\end{abstract}


\begin{IEEEkeywords}
Analog-to-digital conversion, convex optimization, forward-backward splitting (FBS), frequency-selective channels, massive multi-user multiple-input multiple-output (MU-MIMO), maximum a-posteriori (MAP) channel estimation, minimum mean-square error (MMSE)  data detection, orthogonal frequency-division multiplexing (OFDM), quantization.
\end{IEEEkeywords}




\section{Introduction}

\begin{figure*}[t]
\centering
 \includegraphics[width=1.95\columnwidth]{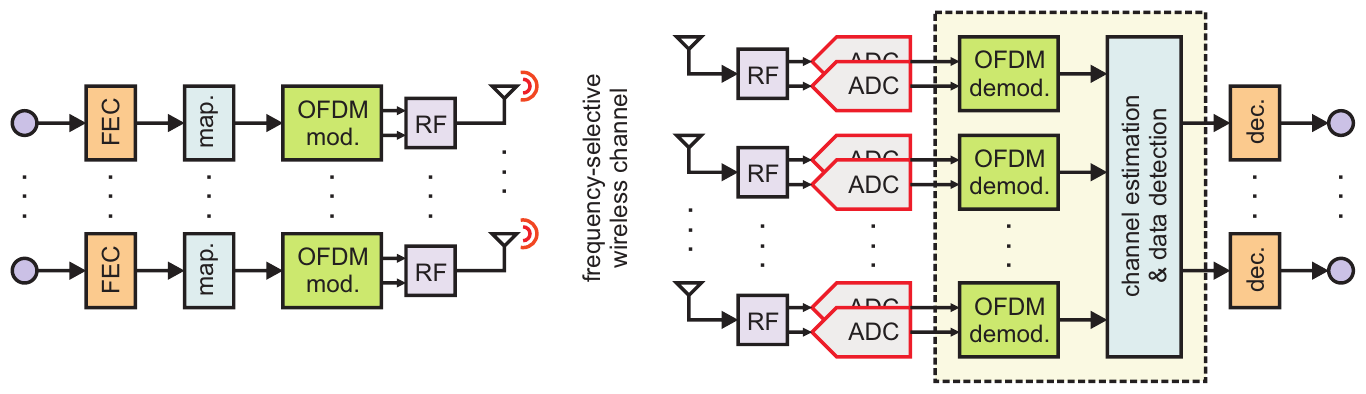}
  \caption{Overview of the considered quantized massive MU-MIMO-OFDM uplink system. Left: $U$ single-antenna user terminals; right: massive MIMO base station with $B\gg U$ antennas and  coarse quantization in the analog-to-digital converters~(ADCs). The proposed system includes OFDM to simplify communication over frequency-selective, wideband channels.}
  \label{fig:overview}
\end{figure*}

We investigate the effects of coarse quantization at the base-station (BS) on the performance of uplink data transmission in a massive multi-user (MU) multiple-input multiple-output (MIMO) system. 
To enable reliable communication over frequency-selective channels, we develop new methods for (optimal) channel estimation and data detection for the case when massive MU-MIMO is combined with orthogonal frequency-division multiplexing (OFDM).
\fref{fig:overview} illustrates the massive MU-MIMO-OFDM uplink transmission system studied in this paper.  
We consider a coded massive MU-MIMO system where $U$ independent single-antenna user terminals  communicate with a BS equipped with $B\gg U$ antennas. 
Our model includes coarse quantization in the analog-to-digital converters (ADCs) at the in-phase and quadrature outputs of each radio-frequency (RF) chain at the BS side. 
The square-dashed box at the BS side encompasses the proposed channel-estimation and data-detection algorithms, which---as we shall see---enable reliable wideband data transmission, even when the ADCs have low precision (e.g., four to six quantization~bits).

\subsection{Benefits of Quantized Massive MIMO}
In conventional multiple-antenna BSs, each RF port is connected to two high-precision ADCs (typically, the in-phase and the quadrature signal components are quantized with resolutions exceeding $10$\,bit). 
Scaling such architectures to massive MIMO with hundreds or thousands active antenna elements would result in prohibitively high power consumption and hardware costs, especially for systems operating in the millimeter-wave frequency bands.
Since the power consumption of an ADC scales roughly exponentially in the number of quantization bits~\cite{murmann14-11a}, reducing the precision of the ADCs provides an effective means to accommodate a massive array of active antenna elements, while---at the same time---keeping the power consumption and system costs within desirable limits.  
Furthermore, the use of low-precision ADCs allows for a relaxation of the quality requirements on the RF circuitry (e.g., low-noise amplifiers, oscillators, and mixers), which enables a further reduction in both the power consumption and  system costs.
The underlying reason is that RF circuitry needs to operate at precision levels ``just above'' the quantization noise floor. 
Apart from power and cost reductions, coarsely quantized massive MIMO also reduces the amount of bits per second that needs to be transferred from the antenna unit (usually located on top of the cell tower) to the baseband processing unit (usually located at the bottom of the cell tower). 
The deployment of low-precision ADCs at the BS mitigates this data-transfer bottleneck.

\subsection{Relevant Prior Art}
Massive MU-MIMO promises to increase the spectral efficiency at reduced signal-processing complexity compared to that of conventional, small-scale MIMO systems,  and is therefore considered to be a key technology for next-generation wireless systems~\cite{Marzetta10,larsson14-02a}. 
Most theoretical results on the performance of  the massive MU-MIMO uplink are for the scenario where the base-station is equipped with infinite-precision ADCs.
The performance impact of low-precision ADCs has  been analyzed only recently.

Bj\"ornsson \emph{et al.}~\cite{bjornson14-11a} modeled the aggregate effect of various residual hardware impairments (including quantization noise) as additive Gaussian noise that is independent of the transmit signal. Based on this model, it was concluded that massive MU-MIMO exhibits a certain degree of robustness against hardware impairments. 
While the underlying Gaussian impairment model was demonstrated to be accurate for a diverse set of residual RF impairments~\cite{Studer_Tx_OFDM,gustavsson14-12a}, this model is a poor match for \emph{coarse} quantization, which causes deterministic distortions that are---by nature---dependent on the transmit signal. 

Risi \emph{et al.}~\cite{RPL14} analyzed the performance of a massive MU-MIMO uplink system under the assumptions that the ADCs have 1-bit resolution, that the users transmit QPSK symbols, that the receiver performs maximum-ratio combining or zero forcing, and that pilot transmission followed by least-squares (LS) channel estimation at the BS is used to acquire channel-state information (CSI) at the receiver.
Their result---albeit limited to QPSK---shows that massive MU-MIMO is resilient against coarse-quantization noise. 
The motivation for using LS channel estimation is that the maximum a-posteriori (MAP) estimator is claimed to have a computational complexity that scales exponentially in the number of active users. 

The analysis in~\cite{RPL14} has   been extended recently to higher-order modulation schemes in~\cite{jacobsson15-06a}. 
There, it was shown that amplitude information about the transmitted signal can be recovered even in the presence of a single-bit quantizer, provided that the number of receive antennas is sufficiently large and the signal-to-noise ratio (SNR)  is not too high.
The observation that noise can help recovering magnitude information under 1-bit quantization has also been made in  the compressive-sensing literature~\cite{davenport14-a,knudson14-04a}. 

Liang and Zhang~\cite{liang15-04a} considered a mixed-ADC massive MU-MIMO architecture, where few high-resolution ADCs are used together with many 1-bit ADCs. 
The high-resolution ADCs allow the receiver to acquire accurate channel-state information (CSI). 
Generalized mutual information was used to characterize the performance of this architecture for the case of Gaussian codebooks and mismatched nearest-neighbor decoding. 
In~\cite{liang15-04a}, it is also noted that dithering prior to quantization may help improving the performance.

In summary, the results in~\cite{bjornson14-11a,RPL14,jacobsson15-06a,liang15-04a} suggest that massive MU-MIMO provides resilience against noise caused by coarse quantization.
However, all these results pertain to massive MU-MIMO systems operating over frequency-flat fading channels. 
Such channels are rarely encountered in modern broadband wireless communication systems, which instead experience a high degree of frequency selectivity.
This paper investigates the impact of coarse ADC quantization on the error-rate performance in systems operating over frequency-selective channels, which is---as we believe---more relevant for next-generation wireless systems.
Although a frequency-selective channel can be converted into a set of independent frequency-flat channels by means of OFDM, such an orthogonalization relies on the assumption that infinite-precision ADCs are used on the time-domain signals (see Section~\ref{sec:simpleOFDM} for the details).

Outside the massive-MIMO literature, the performance of communication systems employing low-precision ADCs has been previously analyzed, mainly in the context of ultra-wideband and millimeter-wave systems. 
The structure of the capacity-achieving input distribution for point-to-point AWGN channels with low-precision ADCs has been characterized in~\cite{singh09-12a}. An extension  to  fading channels in the limiting regime of 1-bit ADCs for the case of both no CSI and of CSI available only  at the receiver was provided in~\cite{mezghani07-06a,mezghani08-07a,mezghani09-06a}.
The case of full CSI at both the transmitter and receiver has been  analyzed recently in~\cite{mo14-10a}.
In the low-SNR regime, it was shown in~\cite{koch13-09a} that a zero-threshold comparator is not optimal for 1-bit ADCs.  
Instead, the capacity-achieving strategy requires the combination of flash signaling~\cite[Def.~2]{verdu02-06a}  with a suitably-chosen quantization threshold.

\revision{The optimization of the ADC levels in single-antenna  communication systems operating over deterministic (i.e., non-fading) frequency-selective channels was tackled in~\cite{narasimha12-07a} using bit error rate as the performance metric, and in~\cite{zeitler12-09a} using the mutual information.}
\revision{Sampling faster than the Nyquist rate may be beneficial when using low-precision ADCs~(see, e.g., \cite{shamai-shitz94-06a,koch10-11a} and~\cite[Sec. VI]{zhang12-02a}). For example, it was shown in~\cite{krone12-05a} that by oversampling at the receiver, one can  reliably transmit 16-QAM symbols over a 1-bit-quantized single-antenna AWGN~channel.}

The strong nonlinearity introduced by low-precision ADCs renders the tasks of MIMO channel estimation and data detection challenging.
Such problems can be cast in the more general framework of parameter estimation in the presence of quantized output measurements~\cite{dabeer06-12a,mezghani10-02a,zeitler12-06a,zymnis2010,boufounos08-03a,plan13-a}.
The specific case of channel estimation has been discussed in~\cite{lok98-08a}, where the maximum-likelihood (ML)  estimate is computed via expectation maximization. 
For the case of 1-bit ADCs combined with time-multiplexed pilot transmission, it was shown recently that the ML estimate has a closed-form expression~\cite{ivrlac07-02a}.
Data detection algorithms have beed discussed, e.g., in~\cite{wang14-04a,mezghani2010belief,mezghani12-03a,choi15-07a}.
All these works, however, deal exclusively with systems operating over frequency-flat  (or narrowband) fading channels.

\subsection{Contributions}

We investigate the effects of coarse quantization on a massive MU-MIMO-OFDM uplink system operating over frequency-selective, wideband channels. 
We develop new channel-estimation and data-detection algorithms, and investigate their performance/complexity trade-offs. Our main contributions are summarized as follows:
\begin{itemize}
\item We show that, under suitable assumptions on the statistics of the fading channel and of the noise, MAP channel estimation in the presence of (coarse) quantization is a convex problem that can be solved exactly using computationally efficient numerical methods. 
\item We formulate the problem of MAP data detection in the presence of quantization and develop a low-complexity minimum-mean square error (MMSE) data-detection algorithm.
\item We develop channel-estimation and data-detection algorithms for two \emph{mismatched} quantization models, which trade error-rate performance for computational complexity. This trade-off is investigated through a comprehensive set of numerical simulations. 
\end{itemize}
Our results demonstrate that massive MIMO enables the use of coarse quantization at the BS, without a significant performance loss and without an increase in computational complexity compared to the case of systems with infinite-precision quantizers.
\revision{Although we focus on the massive MU-MIMO-OFDM uplink,  our algorithms also apply to traditional, small-scale MIMO-OFDM or even single-antenna OFDM~systems.}

\subsection{Notation}
Lowercase and uppercase boldface letters designate column vectors and matrices, respectively. 
For a matrix $\bA$, we indicate its  conjugate transpose by $\bA^H$. The $\ell$th column of the matrix $\bA$ is denoted by~$\bma_\ell$, the entry on the $k$th row and on the $\ell$th column is $A_{k,\ell}$, and the $k$th entry of a vector $\veca$ is~$[\veca]_k=a_k$. 
The $M\times M$ identity matrix is denoted by $\bI_M$ and the $M\times N$ all-zeros matrix by~$\mathbf{0}_{M\times N}$. 
The real and imaginary parts of a complex scalar $a$ are $a^R$ and $a^I$, respectively. 
The multivariate complex-valued circularly-symmetric Gaussian probability density function with  covariance matrix $\bK$ is denoted by~$\setC\setN(\bZero,\bK)$.
\revision{We use $p(\cdot)$ to denote generic probability mass functions, whereas $f(\cdot)$ is reserved for probability density~functions.}

\subsection{Paper Outline}
The rest of the paper is organized as follows. 
\fref{sec:quantizationsec} introduces the quantization model. 
\fref{sec:mainideas} describes the proposed channel-estimation and data-detection algorithms for a simplified single-antenna OFDM system. 
\fref{sec:uplink} deals with the extension of the proposed algorithms to the full-fledged quantized MU-MIMO-OFDM system illustrated in Fig.~\ref{fig:overview}.
\fref{sec:solvers} summarizes the numerical methods used to solve the convex optimization problems for channel estimation and data detection.
\fref{sec:simulation} provides numerical simulation results. 
We conclude in \fref{sec:conclusions}. 


\section{Exact and Mismatched Quantization Models}
\label{sec:quantizationsec}

We now introduce the quantization model considered throughout the paper. We also propose a mismatched quantization model that enables suboptimal receiver algorithms requiring  lower computational complexity. 

\subsection{Quantization of Complex-Valued Data}
\label{sec:quantization}
The in-phase and quadrature components of an RF-chain's outputs are typically converted to the digital domain using a pair of ADCs. 
After a sample-and-hold stage, the analog time-domain sample $s\in\complexset$ is mapped onto a finite-cardinality quantization alphabet according to $q = \setQ(s)$, where $\setQ(\cdot):\complexset \to \setA_c$ is a complex-valued scalar quantizer  and $\setA_c$ is the complex-valued quantization alphabet. 
In what follows, we focus on scalar quantizers that operate independently on the in-phase (real) and on the quadrature (imaginary) part of each sample. With a slight abuse of notation, we frequently apply the quantizer to a vector/matrix, with the understanding that  quantization is applied entry-wise to the vector/matrix.

The real part $s^R$ and the imaginary part $s^I$ of each complex sample $s$ are mapped onto a label from the quantization set $\setA=\{1,\dots,Q\}$ (the same for both real and imaginary part) of cardinality $Q=|\setA|$.  
Hence, we have that $\setA_c=\setA\times\setA$. 
Each label describes the quantization bin in which $s^R$ and $s^I$ fall into.
Consequently, the quantizer output $\setQ(s)$ associated to a complex-valued sample $s=s^R+i s^I$ consists of a pair $(q^R,q^I)\in \setA\times\setA$, or, equivalently, of a complex number $q=q^R+iq^I$.
The labels are determined by comparing $s^R$ and $s^I$ with $Q+1$ quantization bin boundaries $\{b_q\}$:
\begin{align*}
-\infty=b_1<b_2<\cdots<b_{Q}<b_{Q+1}=+\infty.
\end{align*}
Specifically, if $b_m\leq s^R < b_{m+1}$, $m=1,\dots,Q$, then $q^{R}=m$.
The imaginary part of the quantizer output $q^{I}$ is obtained by applying the same rule to $s^I$. 
We use $Q_b=\log_2|\setA|$ to designate the number of quantization bits per real dimension.
The total number of bits per  complex-valued sample is $2Q_b$.

In what follows, we shall focus on the scenario where the thermal noise in the system (which appears before the quantizer) is a stationary memoryless circularly symmetric complex Gaussian process, whose samples have variance $\sigma^2=\No/2$ per real and imaginary part.
With this assumption, we obtain the following nonlinear input-output relation \revision{for the quantizer}:
\begin{align} \label{eq:quantizationmodel}
q = \setQ(z+n).
\end{align}
Here, $z\in\complexset$ is the noiseless, unquantized (time-domain) sample and $n\sim \setC\setN(0,\No)$ is the thermal (receive) noise that is added to the signal of interest prior to quantization.
Throughout the paper,  we make frequent use of the likelihood $p(q\,|\,z)$ 
of a quantization label $q\in\setA_c$  given a noiseless unquantized time-domain sample~$z$.
To compute this quantity, we use the facts that\! \begin{inparaenum}[(i)] 
  \item the real and the imaginary parts of the thermal  noise $n$ are  independent  and 
  \item the complex-valued quantizer $\setQ(\cdot)$ operates independently on the real and the imaginary part. 
\end{inparaenum}  
These two properties imply that
\begin{align} \label{eq:likelihoodfunction}
p(q\,|\,z=z^R+iz^I) = p(q^R\,|\,z^R)p(q^I\,|\,z^I).
\end{align}

We can now compute the probability of observing the real part $q^R$ of the quantization label, given the real part $z^R$  of the noiseless unquantized sample as follows~\cite{zymnis2010,mezghani2010belief,Bellasi13}:
\begin{align}
p(q^R\,|\,z^R) & = p\bigl(\ell(q^R)\leq\,z^R+n^R< u(q^R)\bigr) \nonumber \\
 & = \int_{\ell(q^R)}^{u(q^R)} \frac{1}{\sqrt{2\pi\sigma^2}} \exp\!\left(-\frac{(z^R-\nu)^2}{2\sigma^2}\right)\!\text{d}\nu \nonumber \\
 & =  \Phi\!\left(\frac{u(q^R)-z^R}{\sigma}\right)\! - \Phi\!\left(\frac{\ell(q^R)-z^R}{\sigma}\right)\!.\label{eq:likelihood_real}
\end{align}
Here, $\Phi(a)=\int_{-\infty}^a ({1}/{\sqrt{2\pi}})\exp(-{\nu^2}/{2})\text{d}\nu$ is the cumulative distribution function of a (real-valued) standard normal random variable;  $u(q^R)=b_{(q^R+1)}$ and $\ell(q^R)=b_{(q^R)}$  are the upper and the lower  bin boundary positions associated with the quantized measurement $q^R\in\setA$, respectively. 
The probability of observing the imaginary part $q^I$ of the quantization label, given $z^I$ is obtained analogously as
\begin{align}\label{eq:likelihood_imag}
p(q^I\,|\,z^I) & =  \Phi\!\left(\frac{u(q^I)-z^I}{\sigma}\right) - \Phi\!\left(\frac{\ell(q^I)-z^I}{\sigma}\right)\!.
\end{align}

\subsection{First Mismatched Quantization Model}
\label{sec:mismatchmodel}

The above (exact) quantization model often leads to computationally expensive numerical algorithms that require high arithmetic precision. 
We therefore also consider the case where the receiver algorithms are designed on the basis of \emph{mismatched} quantization models; such models enable the development of simpler and (often) faster algorithms at the cost of a small performance loss.
By ``mismatched quantization model,'' we mean that algorithms for channel estimation and data detection are developed on the basis of a quantized input-output relation that does not match the exact one given in~\eqref{eq:quantizationmodel}.\footnote{The term ``mismatched'' is commonly used in the information-theoretic literature to denote the scenario where the decoder operates according to a specific, possibly suboptimal, rule; see e.g.,~\cite{ganti00-11a} and references therein.}
Obviously, such a mismatch yields an error-rate performance loss.

To describe our first mismatched quantization model (which we abbreviate as ``Mismatch 1''), it is convenient to assign to each quantization label $q$ a complex value $y(q)$ whose real and imaginary parts lie within the quantization bin boundaries $\setS(q)=\bigl[\ell(q^R),u(q^R)\bigr)\times\bigl[\ell(q^I),u(q^I)\bigr)$ associated to $q$.  
Let $f(v)$ denote the probability density function of the unquantized noisy signal $v=z+n$.
Following~\cite{zymnis2010}, we take $y(q)$ as the centroid of $\setS(q)$ under $f(v)$, i.e.,
\begin{align} \label{eq:qmean}
y(q) = \frac{\int_{\setS(q)}vf(v)\text{d}v }{\int_{\setS(q)} f(v)\text{d}v}.
\end{align}
Now we can write
\begin{align} \label{eq:simplemodel}
y(q) = z + n + e
\end{align}
for some quantization error $e$ that depends on the unquantized signal $z+n$.
Note that, so far, we have just provided an alternative description of~\eqref{eq:quantizationmodel}.
We now obtain a mismatched quantization model by assuming that the quantization error $e$ is \emph{independent} of both $z$ and $n$. 
Furthermore, we shall approximate its distribution by a circularly symmetric complex Gaussian distribution with zero mean and variance
\begin{align}\label{eq:qvar}
\gamma^2(q) = \frac{\int_{\setS(q)}|v-y(q)|^2f(v)\text{d}v }{\int_{\setS(q)} f(v)\text{d}v},
\end{align}
which depends on the quantization label $q$.
This approximation has been suggested in~\cite{zymnis2010} in the context of  sparse signal recovery from noisy, quantized measurements.

In this first mismatched quantization model, the likelihood $\tilde{p}(q\,|\,z)$ takes the following form (cf.~\eqref{eq:likelihoodfunction} for the exact model):
\begin{align} \label{eq:simplelikelihoodfunction}
\tilde{p}(q\,|\,z) = \frac{1}{\pi(\No+\gamma^2(q))}\exp\!\left(-\frac{|z-y(q)|^2}{\No+\gamma^2(q)}\right)\!.
\end{align}
To compute \eqref{eq:qmean} and \eqref{eq:qvar}, one can either approximate the probability density function $f(v)$ with a uniform distribution over the set  $\setS(q)$, or---if known---use the distribution of the unquantized signal $v=z+n$.

In summary, the first mismatched quantization model is obtained by assuming (i) 
that the quantization error does not depend on the signal $z$ or the noise $n$,  and
(ii) that the quantization error is Gaussian. 
We shall see that despite the crudeness of these approximations, the first mismatched quantization model enables computationally efficient channel-estimation and data-detection algorithms, which closely approach the error-rate performance of (more complex) algorithms developed for the exact quantization model detailed in \fref{sec:quantization}.

\section{Quantized Single-Input Single-Output OFDM}
\label{sec:mainideas}
We now explain the principles of the channel-estimation and data-detection algorithms proposed in this paper.
In order to avoid that the key features of our algorithms be obfuscated by the unavoidably intricate notation needed to describe a quantized massive MU-MIMO-OFDM uplink system, we first focus on a simple, quantized single-input single-output (SISO) OFDM system.
The full-fledged quantized massive MU-MIMO-OFDM uplink system will be discussed in \fref{sec:uplink}.

\subsection{Quantized SISO-OFDM System Model}
\label{sec:simpleOFDM}
Let the set $\setO$ contain the points of the chosen digital modulation format (e.g., QAM or PSK).
Assume that the transmitter converts a $W$-tone frequency-domain signal $\vecs\in\setO^W$ into the time domain using an inverse discrete Fourier transform (DFT) according to $\vecv=\bF^H\vecs$.
Here, $\bF$ denotes  the $W\times W$ DFT matrix that satisfies $\bF^H\bF=\bI$.
After prepending the cyclic prefix, the signal is transmitted over a frequency-selective channel. The receiver quantizes the baseband time-domain signal and discards the cyclic prefix. 
The resulting time-domain input-output relation is given by 
\begin{align} \label{eq:OFDMexamplechannel}
\vecq = \setQ(\bH\bF^H\vecs+\vecn).
\end{align}
Here, the vector $\vecq\in\setA_c^W$ contains the quantization labels, which are obtained by applying the quantizer $\setQ(\cdot)$ element-wise to the vector $\bH\bF^H\vecs+\vecn$.
The $W\times W$ channel matrix $\bH$ is circulant and contains the samples of the channel's impulse response.
The vector $\vecn\in\complexset^W$ has independent and identically distributed (\iid) circularly symmetric complex Gaussian entries with variance $N_0$.

Since circulant matrices are diagonalized by the Fourier transform, we can rewrite~\eqref{eq:OFDMexamplechannel} as 
\begin{align} \label{eq:OFDMexamplechannel1}
\vecq = \setQ(\bF^H\text{diag}(\bmh)\vecs+\vecn)
\end{align}
where $\text{diag}(\vech)=\bF\bH\bF^H$ stands for a diagonal matrix containing the frequency-domain representation $\vech$ of the channel impulse response on its main diagonal. 
Unfortunately, the nonlinear nature of the quantization operator $\setQ(\cdot)$  implies that it is not possible to convert the time-domain input-output relation~\eqref{eq:OFDMexamplechannel1}  into a \emph{diagonalized} frequency-domain input-output relation, simply by computing a DFT at the receiver.

\subsection{MAP Channel Estimation}
\label{sec:SISOOFDMCHEST}

We now show that pilot-based MAP channel estimation of the channel $\bmh$ in~\eqref{eq:OFDMexamplechannel1} can be formulated as a convex optimization problem. 
Specifically, we assume  a training phase consisting of a single\footnote{The number of  degrees of freedom of frequency-selective channels is typically much smaller than the number of OFDM tones. This property ensures that a single training symbol is sufficient to acquire accurate channel estimates, even in the presence of coarse quantization. Nevertheless, our algorithms can easily be extended to systems that use more than one training symbol.} OFDM symbol, which we model as follows:
\begin{align} \label{eq:channelestimation}
\vecq = \setQ(\bT\vech+\vecn).
\end{align}
Here, $\bT=\bF^H\text{diag}(\bmt)$ where the vector $\vect\in\complexset^W$ contains the pilot symbols (which are known to the receiver).
To obtain~\eqref{eq:channelestimation} from~\eqref{eq:OFDMexamplechannel1}, we set $\bms=\bmt$ and used that $\text{diag}(\bmh)\bmt=\text{diag}(\bmt)\bmh$.
By assuming prior knowledge of the probability density function~$\revision{f}(\bmh)$ of the channel $\bmh$, we can formulate the following MAP channel-estimation problem
\begin{align} \label{eq:simpleMAPchest}
\hat\bmh^\text{MAP} = \argmax_{\tilde\bmh\in\complexset^{W}} \, p(\bmq\,|\,\bT\tilde\bmh)\revision{f}(\tilde\bmh).
\end{align}

We now show that, under mild assumptions on the probability density function $\revision{f}(\bmh)$, the optimization problem~\fref{eq:simpleMAPchest} can be solved \emph{exactly} via convex optimization, despite  the nonlinearity introduced by the quantizer $\setQ(\cdot)$.
To this end, we first take the negative logarithm of the objective function on the right-hand side (RHS) of~\fref{eq:simpleMAPchest}, which allows us to rewrite~\eqref{eq:simpleMAPchest} in the following equivalent form:
\begin{align} \label{eq:simpleMAPchestlog}
\hat\bmh^\text{MAP} = \argmin_{\tilde\bmh\in\complexset^{W}} \,\biggl\{ -\log p(\bmq\,|\,\bT\tilde\bmh)-\log \revision{f}(\tilde\bmh)\biggr\}.
\end{align}
It is now key to realize that the first term $-\log p(\bmq\,|\,\bT\tilde\bmh)$ on the RHS of~\eqref{eq:simpleMAPchestlog} is convex in $\tilde\bmh$.
To show this, we require the following result. 
\begin{thm} \label{thm:convexity}
The negative log-likelihood $-\log p(q\,|\,z)$ in~\eqref{eq:likelihoodfunction} is a smooth, convex function in $z\in\complexset$. 
\end{thm}

\begin{IEEEproof}
We start by noting that  
\begin{align}\label{eq:log_likelihood_function}
  -\log p(q\,|\,z)=-\log p(q^R\,|\,z^R)-\log p(q^I\,|\,z^I).
\end{align}
To prove the theorem, we shall show that both summands on the RHS of~\eqref{eq:log_likelihood_function} are smooth convex functions in the variables $z^R$ and $z^I$. 
Following the line of reasoning in~\cite[Sec.~III-A]{zymnis2010}, we note that the function
\begin{align*}
\Phi(u-c) - \Phi(\ell-c) = \int^{u-c}_{\ell-c} \frac{1}{\sqrt{2\pi}}\exp(-\nu^2/2)\text{d}\nu
\end{align*}
which appears in the expressions for $p(q^R\,|\,z^R)$ and $p(q^I\,|\,z^I)$ in~\eqref{eq:likelihood_real} and~\eqref{eq:likelihood_imag}, respectively,
 is log-concave in the real variable $c$, because it is a convolution between two log-concave functions: the Gaussian pdf and the indicator function of the interval $[\ell,u]$. 
 This concludes the proof of the theorem.
\end{IEEEproof}

Since $\bT\tilde\bmh$ is linear in $\tilde\bmh$, \fref{thm:convexity} implies that the negative log-likelihood 
$-\log p(\bmq\,|\,\bT\tilde\bmh) = \sum_{w=1}^W -\log p(q_w\,|\,[\bT\tilde\bmh]_w)$
is convex in $\tilde\bmh$.
Hence, the MAP channel-estimation problem \fref{eq:simpleMAPchestlog} is convex as long as the negative logarithm of the prior term $-\log \revision{f}(\tilde\bmh)$ is convex.
This property is satisfied by any \emph{log-concave} prior distribution $\revision{f}(\bmh)$ on the channel vector $\bmh$. 
For example, if the entries of $\bmh$ are i.i.d.\ circularly symmetric complex Gaussian with variance $C_0$ (which corresponds to \iid Rayleigh fading), we have 
$-\log \revision{f}(\tilde\bmh) = {\|\tilde\bmh\|_2^2}/{C_0} + \log(\pi C_0)$,
which is a smooth, convex function in $\tilde\bmh$. 

If we assume that no prior knowledge on the probability distribution of $\bmh$ is available, which is equivalent to letting $C_0\to\infty$ in our example, we transform the MAP channel-estimation problem into \revision{an} ML channel-estimation problem, which remains convex. 
The crucial observation that MAP and ML channel estimation in the presence of the quantizer $\setQ(\cdot)$ can be formulated as convex problems implies that efficient numerical methods for channel estimation exist (see \fref{sec:solvers} for the details). 

\subsection{MAP and MMSE Data Detection}
\label{sec:MAPproblemdetection}
To perform data detection from quantized measurements, we first rewrite the input-output relation  \fref{eq:OFDMexamplechannel1} as
$\vecq = \setQ(\widehat\bH\vecs+\vecn)$,
where $\widehat\bH=\bF^H\text{diag}(\hat\bmh)$ and $\hat\bmh$ stands for either the MAP or the ML channel estimate obtained solving the convex optimization problem described in Section~\ref{sec:SISOOFDMCHEST}.

\subsubsection{Hard-Output MAP Detection}
By assuming that prior information on the probability mass function $\revision{p}(\bms)$ of the transmit data vectors $\vecs\in\setO^W$ is available, one could---in principle---minimize the probability that the vector $\bms$ is detected erroneously by solving the following MAP data detection problem:\footnote{\revision{To be precise, the data-detection problem in~\eqref{eq:MAP_hard_data_detection_problem} is the MAP problem only under the assumption of perfect channel state information. 
Since we estimate the channel  in our setup, this detector is actually mismatched. 
This observation holds for all detectors presented in the remainder of this section.}}
\begin{align}\label{eq:MAP_hard_data_detection_problem}
\hat\vecs^\text{MAP} = \argmax_{\tilde\vecs\in\setO^W} \,p(\vecq\,|\,\widehat\bH\tilde\vecs)\revision{p}(\tilde\vecs).
\end{align}
In absence of prior information, one can transform~\eqref{eq:MAP_hard_data_detection_problem} into an ML data detection problem by assuming equally-likely transmit vectors, i.e., that $\revision{p}(\vecs)=|\setO|^{-W}$, $\forall \bms\in\setO^W$.
Since the matrix $\widetilde\bH$ is, in general, not diagonal, this optimization problem is at least as hard as the MAP (or ML) data-detection problem in conventional, unquantized MIMO systems with $W$ spatial streams.
Specifically, solving~\eqref{eq:MAP_hard_data_detection_problem} exactly entails an exhaustive search whose complexity grows exponentially in the number of subcarriers $W$  (see \cite{SJSB11,JO05} for more details). 
Since $W$ can be of the order of hundred or even thousand in modern OFDM systems, solving~\eqref{eq:MAP_hard_data_detection_problem} directly is not feasible and one has to resort to approximate algorithms. 

\subsubsection{Hard-Output and Soft-Output MMSE Detection}
To arrive at a low-complexity data-detection method, we relax the finite-alphabet constraint $\tilde\vecs\in\setO^W$ so that the resulting data-detection problem becomes convex and can be solved efficiently.
We make the common approximation that the entries of $\vecs$ are i.i.d.\ circularly symmetric complex Gaussian random variables with variance~$E_s$ (where $E_s$ stands for the average energy per transmitted symbol); this  allows us to formulate a quantized version of the well-known MMSE equalizer:
\begin{align} \label{eq:QMMSEdetector}
\hat\vecs = \argmin_{\tilde\vecs\in\complexset^W} \biggl\{-\log p(\vecq\,|\,\widehat\bH\tilde\vecs)+\frac{\|\tilde\vecs\|^2_2}{E_s} \biggr\}.
\end{align}
As a consequence of \fref{thm:convexity}, this optimization problem is convex and can be solved efficiently.  
Its solution can either be mapped element-wise onto the closest constellation point as 
$\hat s^\text{MMSE}_w = \argmin_{a\in\setO} |a - \hat s_w|$, $w=1,\ldots,W$, 
to obtain a hard-output MMSE estimate $\hat\bms^\text{MMSE}$, or used to compute soft-information (for a given mapping between bits and constellation symbols) in the form of max-log log-likelihood ratio~(LLR) values \cite{studer2011asic,seethaler2004efficient}:
\begin{align} \label{eq:maxlogLLR}
L_{w,b} = \rho_w \!\min_{a\in\setO^{(0)}_b}|\hat s_w-a|^2- \rho_w\!\min_{a\in\setO^{(1)}_b}|\hat s_w-a|^2.
\end{align}
Here, $\rho_w$ denotes the post-equalization signal-to-interference-and-noise ratio (SINR) on the $w$th subcarrier, and $\setO^{(0)}_b$ and $\setO^{(1)}_b$ refer to the subsets of~$\setO$ consisting of all constellation symbols whose $b$th bit is $0$ and~$1$, respectively.\footnote{The subsets $\setO^{(0)}_b$ and $\setO^{(1)}_b$  depend on the  mapping from  constellation symbols to bits (e.g., a Gray mapping).}  
We note that solving the convex problem \eqref{eq:QMMSEdetector} yields, in general, no information on the SINR $\rho_w$, which makes an exact computation of~\eqref{eq:maxlogLLR} impossible.
Throughout the paper, we approximate~\eqref{eq:maxlogLLR} by setting $\rho_w=1$, unless stated otherwise.
We note that this  approximation was shown in~\cite{WYWDCS2014,yin2014conjugate} to result in near-optimal performance for the case of massive MIMO data detection with max-log channel decoding.

\subsection{Channel Estimation and Data Detection with the First Mismatched Quantization Model}
\label{sec:prettysimplemodel}
We now turn our attention to the first mismatched quantization model introduced in Section~\ref{sec:mismatchmodel} and derive the optimal channel estimator and data detector for the case in which the receiver is designed on the basis of this mismatched model. 
With this mismatched quantization model, the input-output relation~\eqref{eq:OFDMexamplechannel1} is replaced by 
\begin{align} \label{eq:IOrelationofmismatchSISOOFDM}
\vecy(\vecq) = \bF^H\text{diag}(\bmh)\vecs+\vecn+\vece.
\end{align}
In~\eqref{eq:IOrelationofmismatchSISOOFDM}, which is the OFDM generalization of the scalar input-output relation~\eqref{eq:simplemodel}, the noise vector follows a $\setC\setN(0,\No\bI_W)$ distribution, and the entries $\{e_w\}_{w=1}^W$ of the quantization noise vector $\bme$ are distributed according to $\setC\setN(0,\gamma^2(q_w))$. 
The variance $\gamma^2(q_w)$, where $q_w$ denotes the quantization label associated to the received signal on the $w$ subcarrier, is defined similarly to~\eqref{eq:qvar}. 
Proceeding as in the case of unquantized SISO-OFDM systems, we apply a DFT to the time-domain vector $\vecy(\vecq)$ and obtain the following frequency-domain input-output relation:
\begin{align} \label{eq:frequencymodel}
\hat\vecy = \text{diag}(\bmh)\vecs+\hat\vecn+\hat\vece.
\end{align}
Here, $\hat\vecy=\bF\,\vecy(\vecq)$, $\hat\vecn\sim\setC\setN(0,\No\bI_W)$, and $\hat\vece\sim\setC\setN(0,\bK_e)$ where $\bK_e=\bF\boldsymbol\Gamma\bF^H$ is  the covariance matrix of the quantization noise $\hat\vece$, expressed in the frequency domain. 
Finally, $\boldsymbol\Gamma = \text{diag}(\gamma^2(q_1),\ldots,\gamma^2(q_W))$.
We note that the entries of the vector $\hat\vece$ are, in general, correlated.
As we shall discuss next, this has a negative impact on the complexity of MAP channel-estimation and data-detection algorithms.

\subsubsection{MAP Channel Estimation}

Analogously to \fref{eq:simpleMAPchestlog}, MAP channel estimation for the mismatched quantization model~\eqref{eq:frequencymodel} can be formulated as follows:   
\begin{align} \label{eq:MAPCHESTmismatch}
\tilde\bmh^\text{MAP} = \argmin_{\tilde\bmh\in\complexset^{W}} \left\{  \|\bK^{-{1}/{2}}(\hat\vecy - \text{diag}(\bmt)\tilde\vech)\|_2^2
-\log p(\tilde\bmh) \right\}\!.
\end{align}
Here, $\bK=\No\bI_{W} +\bK_e$ stands for the covariance matrix of the Gaussian additive noise plus the frequency-domain quantization noise, and $\bmt\in\complexset^{W}$ denotes the vector containing the pilot symbols used for channel estimation. 
If $p(\tilde\bmh)$ is log-concave, the problem \fref{eq:MAPCHESTmismatch} is also convex and can be solved efficiently using numerical methods. 
In the case where $p(\tilde\vech)$ is Gaussian,~\eqref{eq:MAPCHESTmismatch} reduces to a $W$-dimensional LS problem. 

\subsubsection{MAP and MMSE Data Detection}

Similarly to the MAP channel-estimation problem \fref{eq:MAPCHESTmismatch}, we can formulate the  MAP data-detection problem as follows: 
\begin{align} \label{eq:MAPdetectionmismatch}
\tilde\bms^\text{MAP} = \argmin_{\tilde\bms\in\setO^{W}} \, \biggl\{ \|\bK^{-{1}/{2}}(\hat\vecy - \text{diag}(\hat\bmh)\tilde\vecs)\|_2^2
-\log \revision{p}(\tilde\bms) \biggr\}.
\end{align}
Here, $\hat\bmh$ is the MAP channel estimate, obtained by solving \fref{eq:MAPCHESTmismatch}. 
Since, as already mentioned, $\bK$ is in general not diagonal, solving \fref{eq:MAPdetectionmismatch} entails prohibitive complexity (see the discussion in~\fref{sec:MAPproblemdetection}).  

To reduce the computational complexity of the MAP data-detection problem, we relax the alphabet constraint in \fref{eq:MAPdetectionmismatch} by proceeding as in~\fref{sec:MAPproblemdetection}, and obtain the following mismatched version of the quantized MMSE equalizer in \fref{eq:QMMSEdetector}:
\begin{align} \label{eq:MMSEdetectionmismatch}
\tilde\bms^\text{MMSE} = \argmin_{\tilde\bms\in\complexset^{W}} \, \biggl\{ \|\bK^{-{1}/{2}}(\hat\vecy - \text{diag}(\hat\bmh)\tilde\vecs)\|^2_2 + \frac{\|\tilde\vecs\|_2^2}{E_s} \biggr\}.
\end{align}
The optimization problem in~\eqref{eq:MMSEdetectionmismatch} is a $W$-dimensional LS problem and its solution  can  be used to perform hard-output or soft-output detection, as described in~\fref{sec:MAPproblemdetection}. 

\subsection{Second Mismatched Quantization Model}
\label{sec:supereasymethod}
We now provide a second mismatched quantization model (abbreviated as ``Mismatch 2''), which yields a significant complexity reduction for channel estimation and data detection (cf.~Section~\ref{sec:prettysimplemodel}).  
\revision{The key idea is to replace the noise vector $\bme$ in~\eqref{eq:IOrelationofmismatchSISOOFDM}, whose entries are independent but not identically distributed, with a vector~$\tilde\vece$ with \iid entries. Specifically, we let the entries of~$\tilde\vece$ be $\setC\setN(0,\overline\gamma)$-distributed and set $\overline\gamma=\frac{1}{W}\sum_{w=1}^W\gamma^2(q_w)$, which is the average variance of the entries of the vector $\vece$.}
This \revision{approach} results in the following mismatched quantization model:
\begin{align} \label{eq:mismatchedmodel2}
\tilde{\vecy}(\vecq) = \bF^H\text{diag}(\bmh)\vecs+\vecn+\tilde\vece.
\end{align}
The advantage of this second mismatched quantization model is as follows:
\revision{when we apply the DFT operator to $\tilde{\vecy}(\vecq)$, the total noise remains uncorrelated because the correlation matrix of $\vecn+\tilde\vece$ is a scaled identity (cf.~\eqref{eq:frequencymodel})}.
\revision{This means that the MAP channel-estimation and data-detection problems decouple into $W$ independent one-dimensional problems that can be solved at low computational complexity.}\footnote{This holds provided that the priors are product distributions, i.e., that $\revision{f}(\vech)=\prod_{w=1}^W \revision{f}(h_w)$ and that $\revision{p}(\vecs)=\prod_{w=1}^W \revision{p}(s_w)$.}
In particular, one can deploy  standard OFDM signal-processing techniques that operate tone-wise in the frequency domain.
An immediate benefit of this mismatched quantization model is that we can easily extract the post-equalization SINR values~$\rho_w$, which are required for LLR computation in \fref{eq:maxlogLLR}.


\section{Quantized Massive MU-MIMO-OFDM Uplink}
\label{sec:uplink}

We are now ready to present our quantized massive MU-MIMO-OFDM uplink scheme in full detail.
Although the combination of MU-MIMO and OFDM renders the notation rather involved, the main ideas of our approach follow those discussed in \fref{sec:mainideas} for the quantized SISO-OFDM case.
We start by introducing the system model.
We then derive the MAP channel-estimation and MMSE data-detection algorithms for the exact quantization model. 
Finally, we propose low-complexity algorithm variants based on the mismatched quantization models introduced in Sections~\ref{sec:mismatchmodel} and \ref{sec:supereasymethod}.

\subsection{Quantized MIMO-OFDM System Model}\label{sec:quantized_mimo_ofdm_system_model}
We consider a coded MU-MIMO-OFDM uplink system that employs spatial multiplexing. 
The system model is depicted in \fref{fig:overview} and consists of $U$ independent single-antenna users\footnote{Our framework enables also the use of multi-antenna user equipments. 
In this case, forward error correction and channel decoding can  be performed over the multiple antennas.} communicating with a BS having $B\geq U$ receive antennas. 
We consider a frame-based OFDM transmission, where each OFDM symbol consists of $W$ tones, similar to what is used in IEEE 802.11n~\cite{IEEE11n}.
We furthermore assume that communication is effected in two phases: (i) a training phase consisting of~$U$ OFDM symbols (which enables the training of every channel coefficient at least once) followed by (ii) a data transmission phase consisting of~$D$ OFDM symbols. 
The disjoint sets $\Omega_\text{data}$, $\Omega_\text{pilot}$, and $\Omega_\text{guard}$ contain the indices associated to data tones, pilot tones, and guard (or zero) tones, respectively. 
Clearly, $|\Omega_\text{data}|+|\Omega_\text{pilot}|+|\Omega_\text{guard}|=W$. 
For the sake of simplicity, we assume a sufficiently long cyclic prefix, perfect synchronization, and ignore hardware impairments such as phase noise, I/Q imbalance, and amplifier nonlinearities (see, e.g.,  \cite{schenk2005performance,schenk2008rf,Studer_Tx_OFDM} for more details on hardware impairments in OFDM systems). 

\subsubsection{User Terminal (Transmit Side)}\label{sec:user_terminal}
During the training phase, the data and the pilot tones contain QPSK training symbols that are known to the BS; the guard tones remain unused. 
During the data-transmission phase, each user encodes its information bits using a forward error correction (FEC) channel code and maps the coded bits onto constellation symbols $s_{w}\in\setO$ for $w \in\Omega_\text{data}$ to be transmitted over the data-carrying tones. 
The pilot tones contain  BPSK symbols; in a real system, they can be used to compensate for residual phase noise or frequency offset~\cite{schenk2008rf}. 
The guard tones are still left unused.
For a given OFDM symbol, the frequency-domain transmit vector of each user terminal $u$, which is compactly described by the vector $\vecs_u\in\setO^{W}$, is converted into the time domain using a $W$-point inverse DFT according to $\vecv_u=\bF^H\vecs_u$. After prepending a cyclic prefix of length $P$, the time-domain signals $\vecv_u$ of all users $u=1,\ldots,U$ are transmitted simultaneously and in the same frequency band.

\subsubsection{Base Station (Receive Side)}
Each of the $B$ antennas at the BS receive a noisy mixture of time-domain user data.
Specifically, let $\vecy_b\in\complexset^W$ denote the unquantized time-domain received vector at antenna~$b$,  after removal of the cyclic prefix.  
The time-domain received vector at each antenna~$b=1,\ldots,B$ is quantized according to $\vecq_b=\setQ(\vecy_b)$, and then passed to the channel-estimation/data-detection unit. 
During the training phase, the channel-estimation unit generates estimates $\widehat\bH_w\in\complexset^{B\times U}$, $w=1,\ldots,W$, for all OFDM tones. 
During the data-transmission phase, the data-detection unit generates LLR values  for each coded bit of each user, on the basis of the channel estimates and of the quantized received vectors. 
The resulting (approximate) soft-information is used to perform decoding on a per-user basis. 

\subsubsection{Channel model}
\begin{figure}[t]
\centering
  \includegraphics[width=0.95\columnwidth]{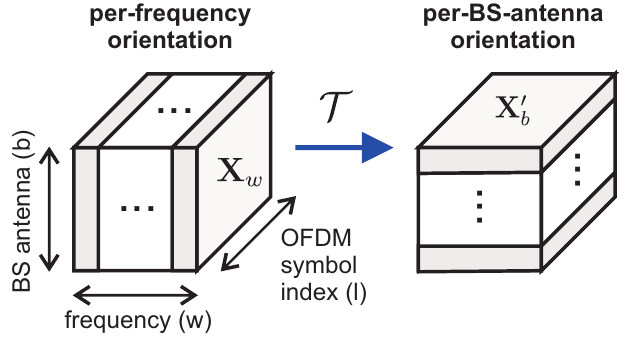}
  \caption{Mapping $\setT$ between per-frequency orientation (left) and per-BS-antenna orientation (right). The data cube (frequency tones $\times$ BS antennas $\times$ OFDM symbols) remains the same but data is associated differently to the matrices $\bX$ and $\bX'$.}
  \label{fig:mapping}
\end{figure}

To simplify the notation, we make frequent use of the mapping~\cite{SL12}  
\begin{align} \label{eq:weirdmapping}
\setT\{ \bX_w\}_{w=1}^W = \{\bX'_b\}_{b=1}^B
\end{align}
between  matrices $\bX_w\in\complexset^{B \times T}$, $w=1,\dots,W$, in \emph{per-frequency} orientation and  matrices $\bX'_b\in\complexset^{W\times T}$, $b=1,\dots,B$, in \emph{per-BS-antenna} orientation. 
Here, $T=U+D$ is the total number of OFDM symbols (during the training and the data-transmission phases). %
Formally, the mapping~\fref{eq:weirdmapping} is defined as $[\bX_w]_{b,\ell}=[\bX'_b]_{w,\ell}$, where  $w=1,\dots,W$, $b=1,\dots,B$, and~$\ell=1,\dots,T$ are the frequency (or time) index, the BS-antenna index, and OFDM-symbol index, respectively. 
To illustrate the effect of the mapping~$\setT$ in~\eqref{eq:weirdmapping},  consider the transmission of a \emph{single} OFDM symbol ($T=1$) consisting of~$W$ subcarriers to a BS equipped with $B$ antennas.
We can either represent the overall received signal as $W$ vectors $\vecx_w\in\setO^B$, each containing $B$  symbols, or, equivalently, as $B$  vectors $\vecx'_b\in\setO^W$, each containing~$W$  symbols. The operator~$\setT$ allows us to  conveniently switch from one representation to the other. 
A pictorial representation of this mapping is shown in \fref{fig:mapping}.

Using the mapping $\setT$ in \fref{eq:weirdmapping} together with the single-antenna OFDM input-output relation \eqref{eq:OFDMexamplechannel1}, we can write the input-output relation of the quantized MU-MIMO-OFDM system as follows:
\begin{align}
\{\bZ_b\}_{b=1}^B & = \setT\{\bH_w\bS_w\}_{w=1}^W \label{eq:freqchannel}\\
\bQ_b & = \setQ(\bF^H\bZ_b+\bN_b), \,\,b=1,\ldots,B.  \notag
\end{align}
Here, the inputs of the channel are the $W$ frequency-domain matrices $\bS_w\in\complexset^{U\times T}$, each containing the data symbols, the pilots, and the guard tones for $T$ OFDM symbols and $U$ users.
The outputs of the channel are the $B$ quantized time-domain matrices $\bQ_b\in\setA_c^{W\times T}$ (one for each BS antenna), containing a $W$-dimensional time-domain signal for each of the $T$ OFDM symbols. 
The frequency-domain channel matrices~$\bH_w\in\complexset^{B\times U}$ in \eqref{eq:freqchannel}, which are assumed to remain constant over the $T$ OFDM symbols,
can be obtained from the time-domain matrices $\bH_t\in\complexset^{B\times U}$, $t=1,\ldots,W$, using the Fourier transform $\{\bH_w\}_{w=1}^W = \setF\{\bH_t\}_{t=1}^W$, defined as follows \cite{bolcskei2002capacity}:
\begin{align}  \label{eq:channelmodel}
\bH_w=  \frac{1}{\sqrt{W}}\sum_{t=1}^{W} \bH_t\exp\!\left(-i2\pi \frac{(t-1)(w-1-W/2)}{W}\right)\!.
\end{align}
Here, the time-domain matrices $\{\bH_t\}_{t=1}^{W}$, of dimension $B\times U$, contain the impulse responses from each user to each BS antenna. In what follows, we assume  that the impulse responses are supported on at most $L<W$ taps, i.e., $\bH_t=\bZero_{B\times U}$ for $t=L+1,\ldots,W$; we furthermore assume that the cyclic-prefix length~$P$ satisfies $L\leq P$.

\subsection{MAP Channel Estimation}
\label{sec:MAPchannelestimation}
We are now ready to formulate MAP channel estimation as a convex optimization problem. As outlined in \fref{sec:mainideas}, we seek estimates $ \widehat\bH_w$ of the channel matrices $\bH_w$, $w=1,\dots, W$, given the quantized measurements $\bQ_b$, $b=1,\ldots,B$, and the known, orthogonal pilot matrices $\bT_w\in\setO^{U\times U}$, where $\omega \in \Omega_\text{pilot}\cup \Omega_{\text{data}}$.
Specifically, our objective is to minimize the negative log-likelihood of the quantized data, given prior information on the channel matrices to be estimated. 
Since it is difficult in practice to acquire accurate prior channel statistics,\footnote{Especially in MU cellular scenarios, where users may experience vastly different path losses.} we assume that the only information known to the receiver about the channel is that its impulse response does not exceed the cyclic prefix length $P$.
This results in the following convex optimization problem:
\begin{align*}
(\text{Q-CHE}) \,\, \left\{\begin{array}{cl} 
\underset{\widetilde\bH_w,w=1,\ldots,W}{\text{minimize}} & \displaystyle -\sum_{b=1}^B \log p(\bQ_b\,|\,\bF^H\bZ_b) \\[0.4cm]
\text{subject to} & \{\bZ_b\}_{b=1}^B = \setT\{\widetilde\bH_w\bT_w\}_{w=1}^W \\[0.1cm]
& \{\widetilde\bH_t\}_{t=1}^W= \setF^{-1}\{\widetilde\bH_w\}_{w=1}^W \\[0.1cm]
& \widetilde\bH_t=\bZero_{B\times U}, \, t=P+1,\ldots,W.
\end{array}\right.
\end{align*}
Here, Q-CHE stands for quantized channel estimation. 
The solution of this convex optimization problem is the set of estimates $\{\widehat\bH_w\}_{w=1}^{W}$  of the MU-MIMO-OFDM channel in the frequency domain. 
The ingredients of the optimization problem (Q-CHE) are as follows. 
The objective function, i.e., the negative log-likelihood of the quantization labels given the noiseless received signal is the MIMO generalization of the objective function given in~\eqref{eq:simpleMAPchestlog} for the SISO-OFDM case.\footnote{Note that, differently from~\eqref{eq:simpleMAPchestlog}, the term corresponding to the prior distribution of the channel is missing from the objective function of (Q-CHE). This is because \emph{a priori} information about the channel distribution is assumed unavailable at the receiver.}
The first constraint simply transforms a per-frequency representation into a per-BS-antenna representation as illustrated in \fref{fig:mapping}; this simplifies the form of the objective function.
The second constraint transforms the channel matrices from the frequency domain to the time domain. 
The third constraint captures that the channel's impulse response  has at most~$P$ nonzero taps. 
Similarly to the SISO-OFDM case, in (Q-CHE) the MAP channel estimation must be performed jointly over all $W$ OFDM tones.
Here, however, we additionally include the constraint that the channel's delay spread does not exceed $P$ taps.
This  enables us to acquire accurate channel estimates using only few OFDM pilot symbols. 
Because of the convexity,  (Q-CHE) can be solved exactly and efficiently using first-order methods (see \fref{sec:solvers} for the details).

\subsection{MMSE Data Detection}
\label{sec:datadetection}
Next, we shall extend the MMSE detector~\fref{eq:QMMSEdetector} to the MU-MIMO case.
The key difference with respect to the SISO-OFDM case is that the resulting convex-optimization problem involves not only all subcarriers, but also all antennas. 
During  the data-transmission phase, the following convex-optimization problem needs to be solved for each OFDM symbol:
\begin{align*}
(\text{Q-DET}) \,\, \left\{\begin{array}{cl}
\underset{\tilde\bms_w, w\in\Omega_\text{data}}{\text{minimize}} & \displaystyle -\sum_{b=1}^B \log p(\bmq_b\,|\,\bF^H\vecz_b) \\[0.4cm]
& \displaystyle + \sum_{w\in\Omega_\text{data}}\!\!\Es^{-1}{\|\bms_w\|_2^2} \\[0.5cm]
\text{subject to} & \{\vecz_b\}_{b=1}^B = \setT\{\widehat\bH_w\vecs_w\}_{w=1}^W\\[0.1cm]
& \bms_w = \bmt_w, w\in\Omega_\text{pilot}.
\end{array}\right.
\end{align*}
Here, Q-DET stands for quantized detection. 
The solution of the optimization problem (Q-DET) are the estimates $\hat\vecs_w\in\Omega_\text{data}$ of the transmitted data symbols; these estimates can then be used to compute hard  or soft estimates as described in \fref{sec:MAPproblemdetection}.
The ingredients of this optimization problem are as follows. 
The objective function is the MIMO generalization of~\eqref{eq:QMMSEdetector} for the SISO-OFDM case.
Specifically, we optimize jointly over the data vectors $\tilde\bms_w$, $w\in\Omega_\text{data}$, transmitted by the $U$ users, and we account for the fact that the base station is equipped with~$B$ receive antennas.
The first constraint in (Q-DET) transforms a per-frequency representation into a per-BS-antenna representation as illustrated in \fref{fig:mapping}.
The second constraint captures that the symbols~$\bmt_w$, $w\in\Omega_\text{pilot}$, transmitted on the pilot tones are known to the receiver (see Section~\ref{sec:quantized_mimo_ofdm_system_model}). 
Similarly to the MAP channel estimation problem (Q-CHE), the problem (Q-DET) can be solved exactly and efficiently  using first-order methods (see \fref{sec:solvers} for the details).

\subsection{Mismatched Quantization Models}

\subsubsection{First Mismatched Quantization Model}\label{sec:first_mismatch_general}
 We now briefly discuss the MAP channel-estimation and data-detection problems for the first mismatched quantization model introduced in \fref{sec:mismatchmodel}. 
In essence,  we only replace the conditional probabilities $p(\cdot|\cdot)$ in (Q-CHE) and (Q-DET) with the mismatched conditional probability  $\tilde{p}(\cdot|\cdot)$ in~\fref{eq:simplelikelihoodfunction}. 
While the dimensionality of the resulting problems remains unchanged, the algorithms that use the mismatched quantizer exhibit improved numerical stability and (often) faster convergence rates at the cost of a small degradation in performance (see \fref{sec:solvers} for the details). 

\subsubsection{Second Mismatched Quantization Model}\label{sec:second_mismatch_general}
For the second mismatched quantization model introduced in \fref{sec:supereasymethod}, the MAP channel estimation and the MMSE data detection problems simplify drastically. 
Specifically, after a DFT operation at the receiver, the channel estimation problem decouples into $U\times B$ independent subproblems, each one involving the estimation of a $W$-dimensional channel vector.
Specifically, each channel estimation subproblem is of the form
\begin{align*}
(\text{MQ-CHE}) \,\, \left\{\begin{array}{ll}
\underset{\tilde\bmh\in\complexset^W}{\text{minimize}} & \displaystyle \sum_{w\in\{\Omega_\text{pilot}\cup\Omega_\text{data}\}} |\hat y_w -\tilde h_w |^2  \\
\text{subject to} & [\bF^H\tilde\bmh]_t=0, \,t=P+1,\ldots,W.
\end{array}\right.
\end{align*}
Here, MQ-CHE stands for mismatched-quantization channel estimation. 
In the above problem, $\hat y_w $, $w\in\{\Omega_\text{pilot}\cup\Omega_\text{pilot}\}$, are the frequency-domain entries  of the matrices $\widetilde\bH_w$, $w\in\{\Omega_\text{pilot}\cup\Omega_\text{pilot}\}$ that correspond to the $u$th user and the $b$th BS antenna, which are estimated using the orthonormal pilot matrices $\bT_w$ (see \fref{sec:MAPchannelestimation}). 
Although (MQ-CHE) has a closed-form solution,\footnote{The problem (MQ-CHE) can be solved by computing a projection matrix of dimension $W\times W$, which involves inverting a matrix  of the same dimension. 
Such a direct approach would not only result in excessive computational complexity but also consume large amounts of memory. } we will use a first-order method to reduce both computational complexity and memory requirements (see \fref{sec:solvers}). In practice, one can also obtain approximate solutions to (MQ-CHE) using more efficient methods (see e.g., \cite{haene2008ofdm}).

With the second mismatched quantization model, MMSE data detection can be performed   \emph{independently} per subcarrier, as it is the case for conventional unquantized MIMO-OFDM systems. Hence, MMSE data detection requires one to solve $|\Omega_\text{data}|$ independent $B\times U$ MMSE data detection problems, which can be done using efficient algorithms. 
Furthermore, the SINR values $\rho_w$ that are necessary for soft-output data detection \fref{eq:maxlogLLR} can  be obtained in this case at virtually no cost; see  \cite{paulraj03,studer2011asic,seethaler2004efficient} for more details on unquantized \revision{(i.e., infinite-precision)} soft-output MMSE data detection. 
%


\section{Efficient Numerical Methods}
\label{sec:solvers}

The problems (Q-CHE), (Q-DET), and  (MQ-CHE) require us to solve large-dimensional convex optimization problems. 
We now summarize a class of algorithms that can be used to solve these problems at low complexity.
We refer the interested reader to~\cite{GSB14} for more details on these algorithms.  

\subsection{Forward-Backward Splitting}
In many fields, including (but not limited to) image processing, machine learning, and spectral clustering, one is interested in solving high-dimensional convex optimization problems of the following form~\cite{BT09}:
\begin{align} \label{eq:generalproblem}
\underset{\bmx\in\complexset^N}{\text{minimize}} \,\, \revision{h}(\bA\bmx)+g(\bmx).
\end{align}
Here, $\bA\in\complexset^{M\times N}$, the function $\revision{h}$ is convex and differentiable, and $g$ is convex, but not necessarily smooth or bounded.
 The generality of the function $g$ prevents the use of simple gradient-descent algorithms. 
  For many problems of interest, however, the so-called \emph{proximal operator} \cite{BT09}
  \begin{align} \label{eq:proximal}
\text{prox}_g(\vecz,\tau) = \argmin_{\vecx} \,\bigl\{\tau g(\vecx) + \textstyle \frac{1}{2}\|\vecx-\vecz\|_2^2\bigr\}
\end{align}
can be computed efficiently.
In this case, one can deploy a computationally efficient first-order method to solve \fref{eq:generalproblem}, generally referred to as forward-backward splitting (FBS). Specifically, FBS performs the following iteration for $k=1,2,\ldots$, until convergence~\cite{BT09}:
\begin{align}\label{eq:FBS_iterations}
\bmx^{\revision{(k+1)}} = \text{prox}_g( \bmx^{\revision{(k)}} - \tau^{\revision{(k)}} \bA^H \nabla \revision{h}(\bA \bmx^{\revision{(k)}}), \tau^{\revision{(k)}}).
\end{align}
FBS requires the computation of the proximal operator~\fref{eq:proximal}, of the gradient $\nabla \revision{h}$ of the smooth function, the computation of matrix-vector products involving  $\bA$ and its adjoint~$\bA^H$, and multiplications by the step size $\tau^{\revision{(k)}}>0$. 
To reduce the complexity of FBS, one should\! \begin{inparaenum}[(i)] \item exploit fast transforms instead of matrix-vector products (if possible) and \item select an appropriate step-size $\tau^{\revision{(k)}}$ in every iteration $k$. 
\end{inparaenum}
While a number of fast FBS implementations have been described in the literature, the numerical results reported in Section~\ref{sec:simulation} use the fast adaptive shrinkage/thresholding algorithm (FASTA)  proposed in~\cite{GSB14}. 

\subsection{Implementation Details}
We shall illustrate next that the optimization problems (Q-CHE), (Q-DET), and  (MQ-CHE) described in Section~\ref{sec:uplink} are of the form given in~\eqref{eq:generalproblem}. 
Indeed, in all these problems the objective function is smooth and convex.
We can therefore identify these objective functions with $\revision{h}$ in~\eqref{eq:generalproblem}.  
The (non-smooth) function~$g$ in~\eqref{eq:generalproblem} can be associated with the affine constraints listed in (Q-CHE), (Q-DET), and  (MQ-CHE). 
To this end, let $\setC$ be the feasible set defined by the affine constraints  of the optimization problems described in Section~\ref{sec:quantized_mimo_ofdm_system_model}, where $\setX_\setC(x)$ is the characteristic function of $\setC$, which we define to be zero if $x\in\setC$ and infinity otherwise. 
We shall set $g(x)=\setX_\setC(x)$, which is a convex (but not smooth or bounded) function. 
Next we detail how to compute the proximal operator~\eqref{eq:proximal} and the gradient of $\revision{h}$, which is needed in~\eqref{eq:FBS_iterations}.

\subsubsection{Gradients} 
For the exact quantization model introduced in \fref{sec:quantization}, the gradient of the negative log-likelihood function $-\log p(\vecq\,|\,\veca)$, for the case of real-valued $\veca$,  is given by \cite{Bellasi13}
\begin{align}\label{eq:gradient_exact}
  [\nabla \revision{h}(\veca)]_i = \frac{\exp\!\left(-\frac{(u(q_i)-a_i)^2}{2\sigma^2}\right)-\exp\!\left(-\frac{(\ell(q_i)-a_i)^2}{2\sigma^2}\right)}{\sqrt{2\pi\sigma^2}\Bigl[\Phi\bigl(\frac{u(q_i)-a_i}{\sigma}\bigr)-\Phi\bigl(\frac{\ell(q_i)-a_i}{\sigma}\bigr)\Bigr]}.
\end{align}
The complex-valued extension of~\eqref{eq:gradient_exact}, which is required  in both (Q-CHE) and (Q-DET) can be readily obtained by computing the gradient separately for real and imaginary parts~\cite{Bellasi13}.
For the mismatched quantization model introduced in \fref{sec:mismatchmodel}, the (complex-valued) gradient of $-\log \tilde{p}(\vecq\,|\,\veca)$ is given by  
\begin{align}\label{eq:gradient_mismatch}
  [\nabla \revision{h}(\veca)]_i = \frac{y(q_i)-a_i}{\No/2+\gamma^2(q_i)}.
\end{align}
A comparison of~\eqref{eq:gradient_exact} and~\eqref{eq:gradient_mismatch} reveals that gradient computation is easier and less computationally intensive for the mismatched quantization model. 

\subsubsection{Proximal Operators} 
As we shall show next, the proximal operators associated with the convex-optimization problems  (Q-CHE), (Q-DET), and  (MQ-CHE) can be evaluated efficiently. 
Specifically, the function $g$ in (Q-CHE) corresponds to  the affine constraint $\widetilde \bH_t=\bZero_{B\times U}$, $t=P+1,\ldots,W$. 
This implies that the proximal operator~\fref{eq:proximal}  involves  transforming the channel matrices from frequency domain to time domain, setting to zero all matrices with index $t=P+1,\ldots,W$ , and transforming the resulting matrices back to the frequency domain. 
By using the fast Fourier transform (FFT), the proximal operator can be evaluated efficiently.  
Similarly, (MQ-CHE) involves the affine constraint $[\bF^H\tilde\bmh]_{t}=0$, $t=P+1,\ldots,W$. 
The computation of the proximal operator requires transforming the frequency-domain vector $\tilde\bmh$ into  time domain, setting to zero all entries with index $t=P+1,\ldots,W$, and transforming the resulting vector back to  frequency domain.
Finally, (Q-DET) involves the affine constraints $\bms_w=\bmt_w$, $w\in\Omega_\text{pilot}$. The proximal operator simply requires one to set  $\bms_w$ equal to $\bmt_w$ for all $w\in\Omega_\text{pilot}$.


\section{Simulation Results}
\label{sec:simulation}

We will now present numerical simulation results for the proposed quantized massive MU-MIMO-OFDM uplink system. 
Due to  space constraints, we shall focus on a selected set of system parameters.\footnote{Our simulation framework will be available for download from GitHub (\url{https://github.com/quantizedmassivemimo/mu_mimo_ofdm/}). The purpose is to enable interested readers to perform their own simulations with different system parameters and also to test alternative algorithms.}

\subsection{Simulation Parameters}
All simulation results  are based on an IEEE 802.11n-like  system~\cite{IEEE11n} with 40\,MHz bandwidth (similarly to the OFDM system parameters considered in \cite{SL12}). 
The number of OFDM subcarriers is $W=128$; guard tones, pilot tones, and data tones are as defined in~\cite{IEEE11n}. Each user encodes its information bits using a (rather weak) rate-$5/6$ convolutional code of constraint length~$7$ with random interleaving. Decoding at the base-station relies on a max-log soft-input  Viterbi decoder.  
\revision{Channel training is performed as described in Section~\ref{sec:user_terminal}.
In the data-transmission phase, the transmitted symbols belong to a 16-QAM constellation.}
The channel matrices are generated according to \fref{eq:channelmodel}, where the entries of the non-zero time-domain matrices $\bH_t$, $t=1,\ldots,L$, are assumed i.i.d.\ circularly complex Gaussian with unit variance. 
We assume that the number of channel taps is $L=4$; the cyclic prefix length is $P=16$. 
The pilot matrices $\bT_w$ are generated from a $U\times U$ orthonormal Hadamard transform matrix, where  the signs of the columns and rows are randomly altered.\footnote{We observed that the choice of the pilot matrices has a noticeable impact on the error-rate performance in systems with coarse quantization. In particular, orthogonalizing the users in time during pilot transmission results in higher channel-estimation errors. A detailed investigation of this phenomenon, \revision{which is similar to the one described in~\cite{jacobsson15-06a} in the context of modulation design for 1-bit massive MIMO systems,} is part of ongoing work.}
\revision{The quantization alphabet $\setA_c$ and the quantization bin boundaries $\{b_q\}$ are set using the Lloyd-Max algorithm~\cite{max60-03a,lloyd82-03a}. 
For a given set of system parameters, the algorithm uses an estimate of the probability density function of the received signal, which is acquired offline.\footnote{\revision{In practice, one can precompute the quantization bin boundaries and store them in a look-up table.}}}
We use the FASTA solver \cite{GSB15software} with  tolerance set to $10^{-6}$ and maximum  number of iterations set to $500$.
The packet error rate (PER), which is averaged over all users, is obtained via Monte--Carlo simulations with 1000 packet transmissions for each simulated point.\footnote{\revision{The trade-off curves are obtained by performing $1000$ Monte-Carlo trials for each SNR point. The irregular behavior of some of the curves is a consequence of the chosen number of trials and the SNR resolution.}}

To characterize the system performance for a given number of quantization bits $Q_b$ (see \fref{sec:quantization}), we compute the so-called ``SNR operating point'' proposed in~\cite{SB10,studer2008soft}, which is the minimum average receive SNR required to achieve $1\%$ PER. 
We consider three receiver architectures. The first one, referred to as ``Quantizer,'' performs MAP channel estimation and MMSE data detection as described in Section~\ref{sec:MAPchannelestimation} and Section~\ref{sec:datadetection}, respectively. 
The second and third one, referred to as ``Mismatch~1'' and ``Mismatch~2,'' respectively, perform their computations as described in Section~\ref{sec:first_mismatch_general} and Section~\ref{sec:second_mismatch_general}, respectively.

\begin{figure*}[t]
\centering
\subfigure[$B=16$ and $U=8$]{\includegraphics[width=0.99\columnwidth]{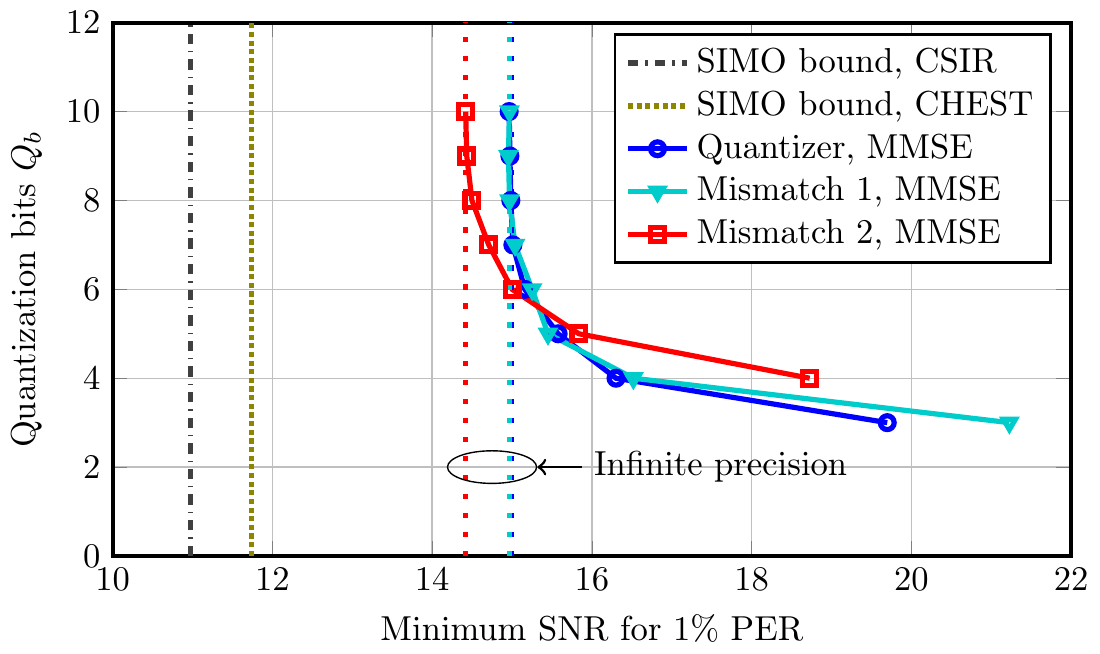}}\hspace{0.2cm}
\subfigure[$B=32$ and $U=8$]{\includegraphics[width=0.99\columnwidth]{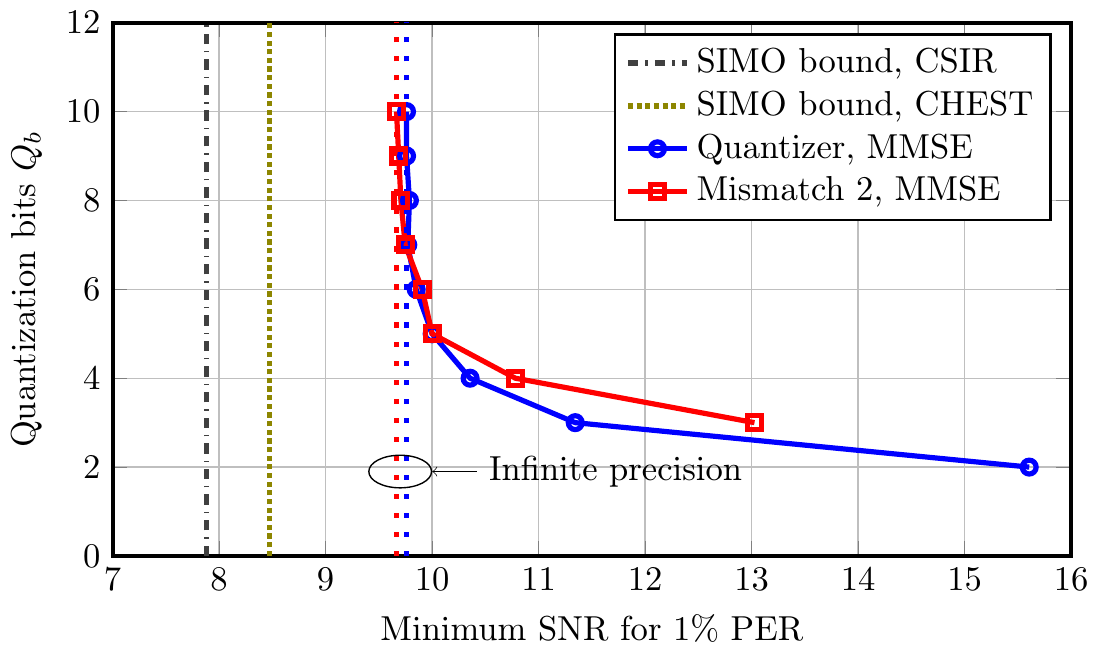}}
\subfigure[$B=64$ and $U=8$]{\includegraphics[width=0.99\columnwidth]{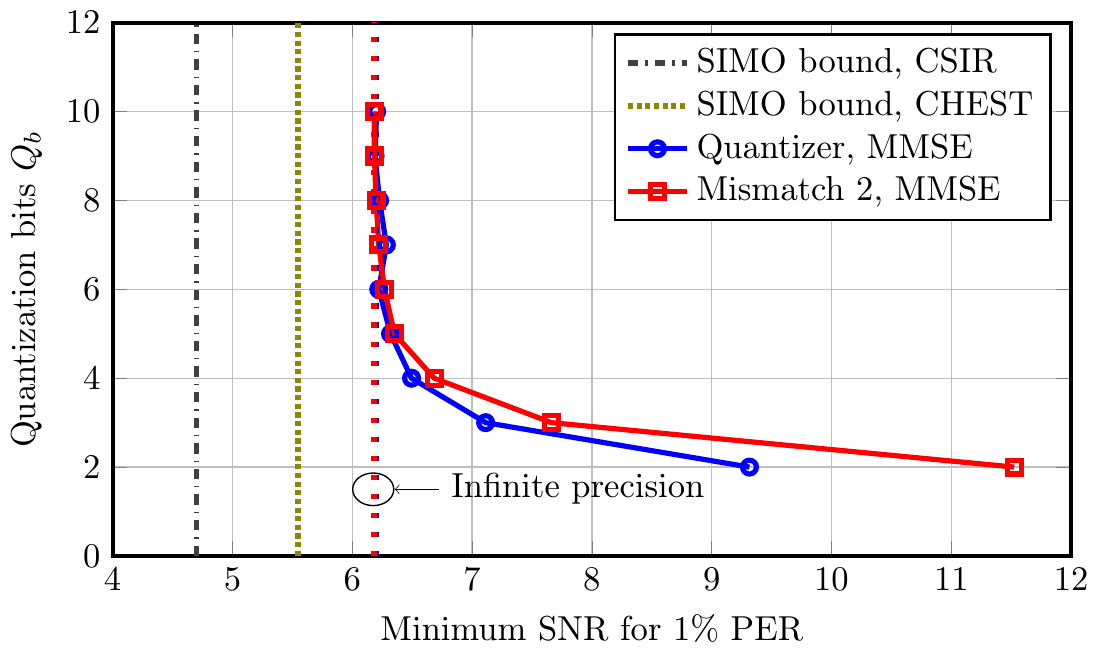}}\hspace{0.2cm}
\subfigure[$B=128$ and $U=8$]{\includegraphics[width=0.99\columnwidth]{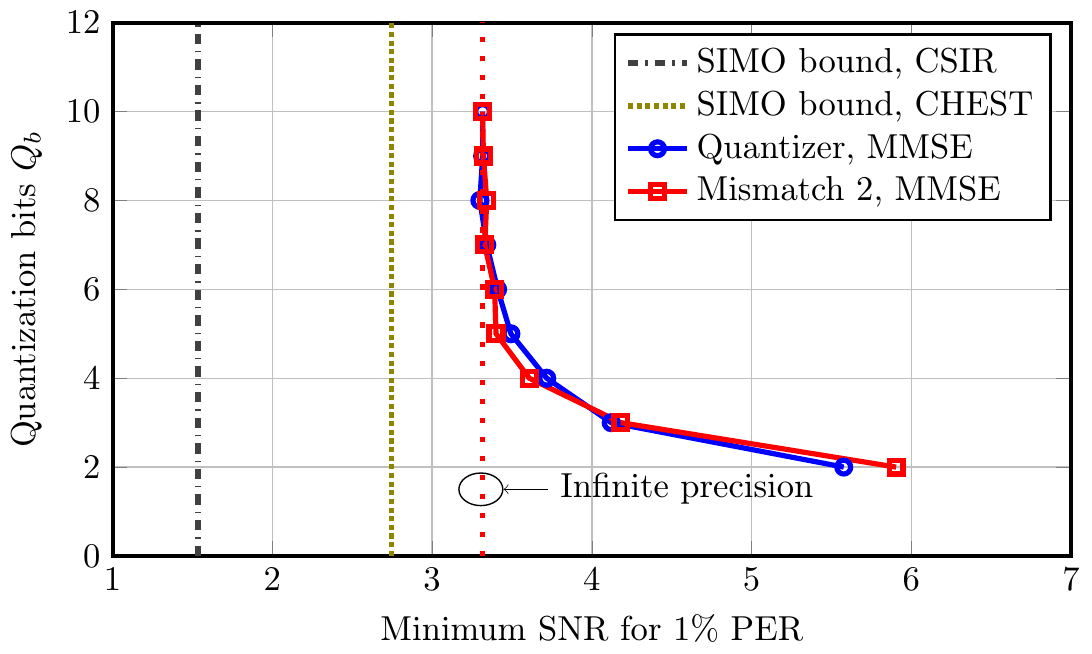}}
  \caption{Trade-off between the SNR operating point and the number of quantization bits $Q_b$ for $8$ users and $\{16,32,64,128\}$ base station antennas. Only a few bits are sufficient to approach the performance of the SIMO lower bound with infinite-precision ADCs. Furthermore, for systems with a large number of BS antennas (e.g., $B=128$), no  sophisticated signal processing is required in the presence of low-precision ADCs, i.e., the ``Mismatch~2'' receiver achieves near-optimal performance.}
\label{fig:tradeoff}
\end{figure*}

\subsection{Results and Discussion}

Fig.~\ref{fig:tradeoff} illustrates the trade-off between the SNR operating point and the number of quantization bits~$Q_b$ for different  BS-antenna-to-user  ratios.
As a reference, we show the performance of an infinite-precision single-input multiple-output (SIMO) system both for the case of perfect CSI at the receiver (CSIR), and for the case of channel estimation (CHEST).
\revision{To obtain these curves, we replaced the negative log-likelihood of the quantized channel output with the one of the channel output before quantization in the objective functions of the iterative algorithms described Section~\ref{sec:uplink}.}

Fig.~\ref{fig:tradeoff}(a) compares the performance/quantization trade-off for the `'Quantizer,'' ``Mismatch 1,''  and ``Mismatch 2'' receiver architectures for a system with $16$ BS antennas and $8$ users (corresponding to a BS-antenna-to-user ratio of $2$). 
We see that ``Quantizer'' outperforms both ``Mismatch~1'' and ``Mismatch~2'' when $3\leq Q_b\leq 5$.
When $Q_b<3$, the system is not able to reach a PER of $1$\%. 
Interestingly, when $Q_b\geq 6$, ``Mismatch~2'' outperforms both ``Quantizer'' and ``Mismatch~1''; this is because in ``Mismatch~2'' it is possible to compute the post-equalization SINR values $\rho_w$ required for soft-output computation in~\fref{eq:maxlogLLR}, whereas in the other receiver architectures, we simply set $\rho_w=1$.

Fig.~\ref{fig:tradeoff}(b) shows that increasing the number of BS antennas to $32$ (corresponding to a BS-antenna-to-user ratio of $4$), enables ``Quantizer'' and ``Mismatch~2'' to achieve a similar SNR operating point when $Q_b\geq 5$. 
This shows that the LLR approximation in \fref{sec:MAPproblemdetection} is accurate for such a BS-antenna-to-user ratio, confirming the observations made in \cite{WYWDCS2014,yin2014conjugate}. 
In this  regime, one should  use the ``Mismatch~2'' receiver architecture (which effectively ignores the presence of a quantizer), as it does not entail a complexity increase compared to conventional algorithms for infinite-precision MU-MIMO-OFDM systems. 
For lower values of $Q_b$, however, ``Quantizer'' significantly outperforms ``Mismatch~2'' (we do not show ``Mismatch~1'' as it exhibits similar performance to ``Quantizer''). 
Also note that with ``Quantizer,'' we are now able to achieve $1$\% PER for only two quantization bits (i.e.,~\mbox{$Q_b=2$}). 

Figs.~\ref{fig:tradeoff}(c) and~\ref{fig:tradeoff}(d) show that by further increasing the BS-antenna-to-user  ratio to $8$ and $16$, respectively, ``Mismatch~2'' delivers the same performance as ``Quantizer.'' In addition, we see that for both receiver architectures, $Q_b=4$ yields an SNR gap to the infinite-precision ADC case of only $0.25$\,dB. Furthermore, the SNR gap to  the SIMO system (with channel estimation, but infinite precision) is less than $1$\,dB.  
All these observations have far reaching consequences for quantized massive MU-MIMO-OFDM systems with BS-antenna-to-user ratio exceeding $8$. In particular, we see that the simple and low-complexity ``Mismatch 2'' receiver architecture  achieves near-optimal performance with only $4$\,bit precision at virtually no complexity increase in baseband processing compared to infinite-precision MU-MIMO-OFDM systems.


\section{Conclusions}
\label{sec:conclusions}

We have studied the performance of a quantized massive MU-MIMO-OFDM uplink operating over frequency-selective, wideband channels. 
We have developed new channel-estimation and data-detection algorithms, and studied the associated performance/quantization trade-offs. 
Our simulation results show that the use of four-bit ADCs is sufficient to achieve near-optimal  performance in massive MU-MIMO-OFDM systems having a BS-antenna-to-user ratio of eight or higher.
Remarkably, this comes at no additional costs in terms of baseband-processing complexity, compared to infinite-precision systems. 
Our results imply that coarse quantization enables low-cost and low-power massive MU-MIMO-OFDM system implementations that achieve near-optimal performance. 
In addition, reducing the precision from $10$\,bit to $4$\,bit yields $2.5\times$ lower ADC output rates, which is  particularly relevant for massive MIMO systems where the antenna elements and baseband processing unit are spatially separated. %

There are many avenues for future work. 
The presence of a quantizer renders synchronization and noise-variance estimation a difficult task; the development of corresponding methods is of paramount importance to enable the practical deployment of quantized massive MU-MIMO-OFDM systems. 
\revision{Extracting post-equalization SINR values for LLR computation directly via convex optimization may yield significant performance improvements; this is an open research problem.}
The application of the algorithms presented in this paper to single-carrier frequency-division multiple access (SC-FDMA) systems with coarse quantization, which is relevant to the 3GPP LTE uplink~\cite{3GPPLTE}, is left for future work. 
\revision{One final open issue is whether the proposed quantized MU-MIMO-OFDM architecture is compatible with systems that use digital filters \emph{after} the ADCs, as typically done in broadband wireless receivers to reduce costs and increase spectrum flexibility.}

 


%
\bibliographystyle{IEEEtran} 

\bibliography{IEEEabrv,confs-jrnls,publishers,studer}

\balance

\end{document}

%% file: vmr-symbols-vecbold.tex
\usepackage{amssymb}
\usepackage{amsfonts}
\usepackage{mathrsfs}
\usepackage{xspace}
\usepackage{bm}
\usepackage{upgreek}

\newcommand{\safemath}[2]{\newcommand{#1}{\ensuremath{#2}\xspace}}

\safemath{\bma}{\mathbf{a}}
\safemath{\bmb}{\mathbf{b}}
\safemath{\bmc}{\mathbf{c}}
\safemath{\bmd}{\mathbf{d}}
\safemath{\bme}{\mathbf{e}}
\safemath{\bmf}{\mathbf{f}}
\safemath{\bmg}{\mathbf{g}}
\safemath{\bmh}{\mathbf{h}}
\safemath{\bmi}{\mathbf{i}}
\safemath{\bmj}{\mathbf{j}}
\safemath{\bmk}{\mathbf{k}}
\safemath{\bml}{\mathbf{l}}
\safemath{\bmm}{\mathbf{m}}
\safemath{\bmn}{\mathbf{n}}
\safemath{\bmo}{\mathbf{o}}
\safemath{\bmp}{\mathbf{p}}
\safemath{\bmq}{\mathbf{q}}
\safemath{\bmr}{\mathbf{r}}
\safemath{\bms}{\mathbf{s}}
\safemath{\bmt}{\mathbf{t}}
\safemath{\bmu}{\mathbf{u}}
\safemath{\bmv}{\mathbf{v}}
\safemath{\bmw}{\mathbf{w}}
\safemath{\bmx}{\mathbf{x}}
\safemath{\bmy}{\mathbf{y}}
\safemath{\bmz}{\mathbf{z}}
\safemath{\bmzero}{\mathbf{0}}
\safemath{\bmone}{\mathbf{1}}

\bmdefine{\biad}{a}
\bmdefine{\bibd}{b}
\bmdefine{\bicd}{c}
\bmdefine{\bidd}{d}
\bmdefine{\bied}{e}
\bmdefine{\bifd}{f}
\bmdefine{\bigd}{g}
\bmdefine{\bihd}{h}
\bmdefine{\biid}{i}
\bmdefine{\bijd}{j}
\bmdefine{\bikd}{k}
\bmdefine{\bild}{l}
\bmdefine{\bimd}{m}
\bmdefine{\bind}{n}
\bmdefine{\biod}{o}
\bmdefine{\bipd}{p}
\bmdefine{\biqd}{q}
\bmdefine{\bird}{r}
\bmdefine{\bisd}{s}
\bmdefine{\bitd}{t}
\bmdefine{\biud}{u}
\bmdefine{\bivd}{v}
\bmdefine{\biwd}{w}
\bmdefine{\bixd}{x}
\bmdefine{\biyd}{y}
\bmdefine{\bizd}{z}

\bmdefine{\bixid}{\xi}
\bmdefine{\bilambdad}{\lambda}
\bmdefine{\bimud}{\mu}
\bmdefine{\bithetad}{\theta}
\bmdefine{\biphid}{\phi}
\bmdefine{\bideltad}{\delta}

\safemath{\bmia}{\biad}
\safemath{\bmib}{\bibd}
\safemath{\bmic}{\bicd}
\safemath{\bmid}{\bidd}
\safemath{\bmie}{\bied}
\safemath{\bmif}{\bifd}
\safemath{\bmig}{\bigd}
\safemath{\bmih}{\bihd}
\safemath{\bmii}{\biid}
\safemath{\bmij}{\bijd}
\safemath{\bmik}{\bikd}
\safemath{\bmil}{\bild}
\safemath{\bmim}{\bimd}
\safemath{\bmin}{\bind}
\safemath{\bmio}{\biod}
\safemath{\bmip}{\bipd}
\safemath{\bmiq}{\biqd}
\safemath{\bmir}{\bird}
\safemath{\bmis}{\bisd}
\safemath{\bmit}{\bitd}
\safemath{\bmiu}{\biud}
\safemath{\bmiv}{\bivd}
\safemath{\bmiw}{\biwd}
\safemath{\bmix}{\bixd}
\safemath{\bmiy}{\biyd}
\safemath{\bmiz}{\bizd}

\safemath{\bmxi}{\bixid}
\safemath{\bmlambda}{\bilambdad}
\safemath{\bmmu}{\bimud}
\safemath{\bmtheta}{\bithetad}
\safemath{\bmphi}{\biphid}
\safemath{\bmdelta}{\bideltad}

\safemath{\bA}{\mathbf{A}}
\safemath{\bB}{\mathbf{B}}
\safemath{\bC}{\mathbf{C}}
\safemath{\bD}{\mathbf{D}}
\safemath{\bE}{\mathbf{E}}
\safemath{\bF}{\mathbf{F}}
\safemath{\bG}{\mathbf{G}}
\safemath{\bH}{\mathbf{H}}
\safemath{\bI}{\mathbf{I}}
\safemath{\bJ}{\mathbf{J}}
\safemath{\bK}{\mathbf{K}}
\safemath{\bL}{\mathbf{L}}
\safemath{\bM}{\mathbf{M}}
\safemath{\bN}{\mathbf{N}}
\safemath{\bO}{\mathbf{O}}
\safemath{\bP}{\mathbf{P}}
\safemath{\bQ}{\mathbf{Q}}
\safemath{\bR}{\mathbf{R}}
\safemath{\bS}{\mathbf{S}}
\safemath{\bT}{\mathbf{T}}
\safemath{\bU}{\mathbf{U}}
\safemath{\bV}{\mathbf{V}}
\safemath{\bW}{\mathbf{W}}
\safemath{\bX}{\mathbf{X}}
\safemath{\bY}{\mathbf{Y}}
\safemath{\bZ}{\mathbf{Z}}

\safemath{\bZero}{\mathbf{0}}
\safemath{\bOne}{\mathbf{1}}
\safemath{\bDelta}{\mathbf{\Delta}}
\safemath{\bLambda}{\mathbf{\UpLambda}}
\safemath{\bPhi}{\mathbf{\Upphi}}
\safemath{\bSigma}{\mathbf{\Upsigma}}
\safemath{\bOmega}{\mathbf{\Upomega}}
\safemath{\bTheta}{\mathbf{\Uptheta}}

\bmdefine{\biAd}{A}
\bmdefine{\biBd}{B}
\bmdefine{\biCd}{C}
\bmdefine{\biDd}{D}
\bmdefine{\biEd}{E}
\bmdefine{\biFd}{F}
\bmdefine{\biGd}{G}
\bmdefine{\biHd}{H}
\bmdefine{\biId}{I}
\bmdefine{\biJd}{J}
\bmdefine{\biKd}{K}
\bmdefine{\biLd}{L}
\bmdefine{\biMd}{M}
\bmdefine{\biOd}{N}
\bmdefine{\biPd}{O}
\bmdefine{\biQd}{P}
\bmdefine{\biRd}{R}
\bmdefine{\biSd}{S}
\bmdefine{\biTd}{T}
\bmdefine{\biUd}{U}
\bmdefine{\biVd}{V}
\bmdefine{\biWd}{W}
\bmdefine{\biXd}{X}
\bmdefine{\biYd}{Y}
\bmdefine{\biZd}{Z}

\bmdefine{\biDelta}{\Delta}
\bmdefine{\biLambda}{\Lambda}
\bmdefine{\biPhi}{\Phi}
\bmdefine{\biSigma}{\Sigma}
\bmdefine{\biOmega}{\Omega}
\bmdefine{\biTheta}{\Theta}

\safemath{\bimA}{\biAd}
\safemath{\bimB}{\biBd}
\safemath{\bimC}{\biCd}
\safemath{\bimD}{\biDd}
\safemath{\bimE}{\biEd}
\safemath{\bimF}{\biFd}
\safemath{\bimG}{\biGd}
\safemath{\bimH}{\biHd}
\safemath{\bimI}{\biId}
\safemath{\bimJ}{\biJd}
\safemath{\bimK}{\biKd}
\safemath{\bimL}{\biLd}
\safemath{\bimM}{\biMd}
\safemath{\bimN}{\biNd}
\safemath{\bimO}{\biOd}
\safemath{\bimP}{\biPd}
\safemath{\bimQ}{\biQd}
\safemath{\bimR}{\biRd}
\safemath{\bimS}{\biSd}
\safemath{\bimT}{\biTd}
\safemath{\bimU}{\biUd}
\safemath{\bimV}{\biVd}
\safemath{\bimW}{\biWd}
\safemath{\bimX}{\biXd}
\safemath{\bimY}{\biYd}
\safemath{\bimZ}{\biZd}

\safemath{\bimDelta}{\biDelta}
\safemath{\bimLambda}{\biLambda}
\safemath{\bimPhi}{\biPhi}
\safemath{\bimSigma}{\biSigma}
\safemath{\bimOmega}{\biOmega}
\safemath{\bimTheta}{\biTheta}

\safemath{\setA}{\mathcal{A}}
\safemath{\setB}{\mathcal{B}}
\safemath{\setC}{\mathcal{C}}
\safemath{\setD}{\mathcal{D}}
\safemath{\setE}{\mathcal{E}}
\safemath{\setF}{\mathcal{F}}
\safemath{\setG}{\mathcal{G}}
\safemath{\setH}{\mathcal{H}}
\safemath{\setI}{\mathcal{I}}
\safemath{\setJ}{\mathcal{J}}
\safemath{\setK}{\mathcal{K}}
\safemath{\setL}{\mathcal{L}}
\safemath{\setM}{\mathcal{M}}
\safemath{\setN}{\mathcal{N}}
\safemath{\setO}{\mathcal{O}}
\safemath{\setP}{\mathcal{P}}
\safemath{\setQ}{\mathcal{Q}}
\safemath{\setR}{\mathcal{R}}
\safemath{\setS}{\mathcal{S}}
\safemath{\setT}{\mathcal{T}}
\safemath{\setU}{\mathcal{U}}
\safemath{\setV}{\mathcal{V}}
\safemath{\setW}{\mathcal{W}}
\safemath{\setX}{\mathcal{X}}
\safemath{\setY}{\mathcal{Y}}
\safemath{\setZ}{\mathcal{Z}}
\safemath{\emptySet}{\varnothing}

\safemath{\colA}{\mathscr{A}}
\safemath{\colB}{\mathscr{B}}
\safemath{\colC}{\mathscr{C}}
\safemath{\colD}{\mathscr{D}}
\safemath{\colE}{\mathscr{E}}
\safemath{\colF}{\mathscr{F}}
\safemath{\colG}{\mathscr{G}}
\safemath{\colH}{\mathscr{H}}
\safemath{\colI}{\mathscr{I}}
\safemath{\colJ}{\mathscr{J}}
\safemath{\colK}{\mathscr{K}}
\safemath{\colL}{\mathscr{L}}
\safemath{\colM}{\mathscr{M}}
\safemath{\colN}{\mathscr{N}}
\safemath{\colO}{\mathscr{O}}
\safemath{\colP}{\mathscr{P}}
\safemath{\colQ}{\mathscr{Q}}
\safemath{\colR}{\mathscr{R}}
\safemath{\colS}{\mathscr{S}}
\safemath{\colT}{\mathscr{T}}
\safemath{\colU}{\mathscr{U}}
\safemath{\colV}{\mathscr{V}}
\safemath{\colW}{\mathscr{W}}
\safemath{\colX}{\mathscr{X}}
\safemath{\colY}{\mathscr{Y}}
\safemath{\colZ}{\mathscr{Z}}

\safemath{\opA}{\mathbb{A}}
\safemath{\opB}{\mathbb{B}}
\safemath{\opC}{\mathbb{C}}
\safemath{\opD}{\mathbb{D}}
\safemath{\opE}{\mathbb{E}}
\safemath{\opF}{\mathbb{F}}
\safemath{\opG}{\mathbb{G}}
\safemath{\opH}{\mathbb{H}}
\safemath{\opI}{\mathbb{I}}
\safemath{\opJ}{\mathbb{J}}
\safemath{\opK}{\mathbb{K}}
\safemath{\opL}{\mathbb{L}}
\safemath{\opM}{\mathbb{M}}
\safemath{\opN}{\mathbb{N}}
\safemath{\opO}{\mathbb{O}}
\safemath{\opP}{\mathbb{P}}
\safemath{\opQ}{\mathbb{Q}}
\safemath{\opR}{\mathbb{R}}
\safemath{\opS}{\mathbb{S}}
\safemath{\opT}{\mathbb{T}}
\safemath{\opU}{\mathbb{U}}
\safemath{\opV}{\mathbb{V}}
\safemath{\opW}{\mathbb{W}}
\safemath{\opX}{\mathbb{X}}
\safemath{\opY}{\mathbb{Y}}
\safemath{\opZ}{\mathbb{Z}}
\safemath{\opZero}{\mathbb{O}}
\safemath{\identityop}{\opI}

\safemath{\veca}{\bma}
\safemath{\vecb}{\bmb}
\safemath{\vecc}{\bmc}
\safemath{\vecd}{\bmd}
\safemath{\vece}{\bme}
\safemath{\vecf}{\bmf}
\safemath{\vecg}{\bmg}
\safemath{\vech}{\bmh}
\safemath{\veci}{\bmi}
\safemath{\vecj}{\bmj}
\safemath{\veck}{\bmk}
\safemath{\vecl}{\bml}
\safemath{\vecm}{\bmm}
\safemath{\vecn}{\bmn}
\safemath{\veco}{\bmo}
\safemath{\vecp}{\bmp}
\safemath{\vecq}{\bmq}
\safemath{\vecr}{\bmr}
\safemath{\vecs}{\bms}
\safemath{\vect}{\bmt}
\safemath{\vecu}{\bmu}
\safemath{\vecv}{\bmv}
\safemath{\vecw}{\bmw}
\safemath{\vecx}{\bmx}
\safemath{\vecy}{\bmy}
\safemath{\vecz}{\bmz}

\safemath{\veczero}{\bmzero}
\safemath{\vecone}{\bmone}
\safemath{\vecxi}{\bmxi}
\safemath{\veclambda}{\bmlambda}
\safemath{\vecmu}{\bmmu}
\safemath{\vectheta}{\bmtheta}
\safemath{\vecphi}{\bmphi}
\safemath{\vecdelta}{\bmdelta}

\safemath{\matA}{\bA}
\safemath{\matB}{\bB}
\safemath{\matC}{\bC}
\safemath{\matD}{\bD}
\safemath{\matE}{\bE}
\safemath{\matF}{\bF}
\safemath{\matG}{\bG}
\safemath{\matH}{\bH}
\safemath{\matI}{\bI}
\safemath{\matJ}{\bJ}
\safemath{\matK}{\bK}
\safemath{\matL}{\bL}
\safemath{\matM}{\bM}
\safemath{\matN}{\bN}
\safemath{\matO}{\bO}
\safemath{\matP}{\bP}
\safemath{\matQ}{\bQ}
\safemath{\matR}{\bR}
\safemath{\matS}{\bS}
\safemath{\matT}{\bT}
\safemath{\matU}{\bU}
\safemath{\matV}{\bV}
\safemath{\matW}{\bW}
\safemath{\matX}{\bX}
\safemath{\matY}{\bY}
\safemath{\matZ}{\bZ}
\safemath{\matzero}{\bmzero}

\safemath{\matDelta}{\bDelta}
\safemath{\matLambda}{\bLambda}
\safemath{\matPhi}{\bPhi}
\safemath{\matSigma}{\bSigma}
\safemath{\matOmega}{\bOmega}
\safemath{\matTheta}{\bTheta}

\safemath{\matidentity}{\matI}
\safemath{\matone}{\matO}

\safemath{\rnda}{A}
\safemath{\rndb}{B}
\safemath{\rndc}{C}
\safemath{\rndd}{D}
\safemath{\rnde}{E}
\safemath{\rndf}{F}
\safemath{\rndg}{G}
\safemath{\rndh}{H}
\safemath{\rndi}{I}
\safemath{\rndj}{J}
\safemath{\rndk}{K}
\safemath{\rndl}{L}
\safemath{\rndm}{M}
\safemath{\rndn}{N}
\safemath{\rndo}{O}
\safemath{\rndp}{P}
\safemath{\rndq}{Q}
\safemath{\rndr}{R}
\safemath{\rnds}{S}
\safemath{\rndt}{T}
\safemath{\rndu}{U}
\safemath{\rndv}{V}
\safemath{\rndw}{W}
\safemath{\rndx}{X}
\safemath{\rndy}{Y}
\safemath{\rndz}{Z}

\safemath{\rveca}{\bimA}
\safemath{\rvecb}{\bimB}
\safemath{\rvecc}{\bimC}
\safemath{\rvecd}{\bimD}
\safemath{\rvece}{\bimE}
\safemath{\rvecf}{\bimF}
\safemath{\rvecg}{\bimG}
\safemath{\rvech}{\bimH}
\safemath{\rveci}{\bimI}
\safemath{\rvecj}{\bimJ}
\safemath{\rveck}{\bimK}
\safemath{\rvecl}{\bimL}
\safemath{\rvecm}{\bimM}
\safemath{\rvecn}{\bimN}
\safemath{\rveco}{\bomO}
\safemath{\rvecp}{\bimP}
\safemath{\rvecq}{\bimQ}
\safemath{\rvecr}{\bimR}
\safemath{\rvecs}{\bimS}
\safemath{\rvect}{\bimT}
\safemath{\rvecu}{\bimU}
\safemath{\rvecv}{\bimV}
\safemath{\rvecw}{\bimW}
\safemath{\rvecx}{\bimX}
\safemath{\rvecy}{\bimY}
\safemath{\rvecz}{\bimZ}

\safemath{\rvecxi}{\bmxi}
\safemath{\rveclambda}{\bmlambda}
\safemath{\rvecmu}{\bmmu}
\safemath{\rvectheta}{\bmtheta}
\safemath{\rvecphi}{\bmphi}

\safemath{\rmatA}{\bimA}
\safemath{\rmatB}{\bimB}
\safemath{\rmatC}{\bimC}
\safemath{\rmatD}{\bimD}
\safemath{\rmatE}{\bimE}
\safemath{\rmatF}{\bimF}
\safemath{\rmatG}{\bimG}
\safemath{\rmatH}{\bimH}
\safemath{\rmatI}{\bimI}
\safemath{\rmatJ}{\bimJ}
\safemath{\rmatK}{\bimK}
\safemath{\rmatL}{\bimL}
\safemath{\rmatM}{\bimM}
\safemath{\rmatN}{\bimN}
\safemath{\rmatO}{\bimO}
\safemath{\rmatP}{\bimP}
\safemath{\rmatQ}{\bimQ}
\safemath{\rmatR}{\bimR}
\safemath{\rmatS}{\bimS}
\safemath{\rmatT}{\bimT}
\safemath{\rmatU}{\bimU}
\safemath{\rmatV}{\bimV}
\safemath{\rmatW}{\bimW}
\safemath{\rmatX}{\bimX}
\safemath{\rmatY}{\bimY}
\safemath{\rmatZ}{\bimZ}

\safemath{\rmatDelta}{\bimDelta}
\safemath{\rmatLambda}{\bimLambda}
\safemath{\rmatPhi}{\bimPhi}
\safemath{\rmatSigma}{\bimSigma}
\safemath{\rmatOmega}{\bimOmega}
\safemath{\rmatTheta}{\bimTheta}

%% file: standard-macros.tex
\usepackage{amssymb}
\usepackage{amsfonts}
\usepackage{mathrsfs}
\usepackage{xspace}
\usepackage{bm}
\usepackage{fancyref}
\usepackage{textcomp}

\usepackage{multirow}
\usepackage{stmaryrd}

\newenvironment{textbmatrix}{	\setlength{\arraycolsep}{2.5pt}%
								\big[\begin{matrix}}{\end{matrix}\big]%
								\raisebox{0.08ex}{\vphantom{M}}}

\def\be{\begin{equation}}
\def\ee{\end{equation}}
\def\een{\nonumber \end{equation}}
\def\mat{\begin{bmatrix}}
\def\emat{\end{bmatrix}}
\def\btm{\begin{textbmatrix}}
\def\etm{\end{textbmatrix}}

\def\ba#1\ea{\begin{align}#1\end{align}}
\def\bas#1\eas{\begin{align*}#1\end{align*}}
\def\bs#1\es{\begin{split}#1\end{split}} 
\def\bg#1\eg{\begin{gather}#1\end{gather}}
\def\bml#1\eml{\begin{multline}#1\end{multline}}
\def\bi#1\ei{\begin{itemize}#1\end{itemize}}

\DeclareMathOperator*{\argmin}{arg\;min}		
\DeclareMathOperator*{\argmax}{arg\;max}		

\safemath{\dirac}{\delta}					
\safemath{\krond}{\dirac}					

\safemath{\upto}{\uparrow}
\safemath{\downto}{\downarrow}
\safemath{\iu}{j}							
\safemath{\ev}{\lambda}						
\safemath{\hilseqspace}{l^{2}}				
\newcommand{\banachfunspace}[1]{\setL^{#1}}	
\safemath{\hilfunspace}{\banachfunspace{2}}	

\safemath{\SNR}{\textsf{SNR}} 				
\safemath{\PAR}{\textsf{PAR}} 				
\safemath{\No}{N_0}							
\safemath{\Es}{E_s}							
\safemath{\Eb}{E_b}							
\safemath{\EbNo}{\frac{\Eb}{\No}}
\safemath{\EsNo}{\frac{\Es}{\No}}

\DeclareMathOperator{\CHop}{\ensuremath{\opH}} 
\safemath{\tvir}{\rndh_{\CHop}}				
\safemath{\tvtf}{\rndl_{\CHop}}				
\safemath{\spf}{\rnds_{\CHop}}				
\safemath{\bff}{H_{\CHop}}					

\safemath{\ircf}{r_{h}}						
\safemath{\tftvcf}{r_{s}}					
\safemath{\tfcf}{r_{l}}						
\safemath{\bfcf}{r_{H}}						

\safemath{\tcorr}{c_h}						
\safemath{\scf}{c_{s}}						
\safemath{\tfcorr}{c_{l}}					
\safemath{\fcorr}{c_{H}}						

\safemath{\mi}{I}							
\safemath{\capacity}{C}						

\newcommand{\iid}{i.i.d.\@\xspace}

\safemath{\normal}{\mathcal{N}}			
\safemath{\jpg}{\mathcal{CN}}			
\safemath{\mchain}{\leftrightarrow}		

\safemath{\dB}{\,\mathrm{dB}}
\safemath{\dBm}{\,\mathrm{dBm}}
\safemath{\Hz}{\,\mathrm{Hz}}
\safemath{\kHz}{\,\mathrm{kHz}}
\safemath{\MHz}{\,\mathrm{MHz}}
\safemath{\GHz}{\,\mathrm{GHz}}
\safemath{\s}{\,\mathrm{s}}
\safemath{\ms}{\,\mathrm{ms}}
\safemath{\mus}{\,\mathrm{\text{\textmu}s}}
\safemath{\ns}{\,\mathrm{ns}}
\safemath{\ps}{\,\mathrm{ps}}
\safemath{\meter}{\,\mathrm{m}}
\safemath{\mm}{\,\mathrm{mm}}
\safemath{\cm}{\,\mathrm{cm}}
\safemath{\m}{\,\mathrm{m}}
\safemath{\W}{\,\mathrm{W}}
\safemath{\mW}{\, \mathrm{mW}}
\safemath{\J}{\,\mathrm{J}}
\safemath{\K}{\,\mathrm{K}}
\safemath{\bit}{\,\mathrm{bit}}
\safemath{\nat}{\,\mathrm{nat}}

\safemath{\define}{\triangleq}			

\safemath{\equivalent}{\sim}
\safemath{\distas}{\sim}					
\safemath{\sdiff}{\Delta}				

\safemath{\reals}{\mathbb{R}}
\safemath{\positivereals}{\reals_{+}}
\safemath{\integers}{\mathbb{Z}}
\safemath{\posint}{\integers_{+}}
\safemath{\naturals}{\mathbb{N}}
\safemath{\posnaturals}{\naturals_{+}}
\safemath{\complexset}{\mathbb{C}}
\safemath{\rationals}{\mathbb{Q}}

\newcommand*{\fancyrefapplabelprefix}{app}		
\newcommand*{\fancyrefthmlabelprefix}{thm}		
\newcommand*{\fancyreflemlabelprefix}{lem}		
\newcommand*{\fancyrefcorlabelprefix}{cor}		
\newcommand*{\fancyrefdeflabelprefix}{def}		
\newcommand*{\fancyrefproplabelprefix}{prop}		
\newcommand*{\fancyrefobslabelprefix}{obs}		
\newcommand*{\fancyrefexmpllabelprefix}{exmpl}
\newcommand*{\fancyrefalglabelprefix}{alg}		

\frefformat{vario}{\fancyrefseclabelprefix}{Section~#1}
\frefformat{vario}{\fancyrefthmlabelprefix}{Theorem~#1}
\frefformat{vario}{\fancyreflemlabelprefix}{Lemma~#1}
\frefformat{vario}{\fancyrefcorlabelprefix}{Corollary~#1}
\frefformat{vario}{\fancyrefdeflabelprefix}{Definition~#1}
\frefformat{vario}{\fancyrefobslabelprefix}{Observation~#1}
\frefformat{vario}{\fancyreffiglabelprefix}{Fig.~#1}
\frefformat{vario}{\fancyrefapplabelprefix}{Appendix~#1} 
\frefformat{vario}{\fancyrefeqlabelprefix}{(#1)}
\frefformat{vario}{\fancyrefproplabelprefix}{Proposition~#1}
\frefformat{vario}{\fancyrefexmpllabelprefix}{Example~#1}
\frefformat{vario}{\fancyrefalglabelprefix}{Algorithm~#1}

%% file: defs.tex
\newtheorem{thm}{Theorem}

\safemath{\dictab}{[\,\dicta\,\,\dictb\,]}

\safemath{\ysig}{\bmy}
\safemath{\ysighat}{\hat{\ysig}}
\safemath{\ysigdim}{M}
\safemath{\xsig}{\bmx}
\safemath{\xsigdim}{N}
\safemath{\nx}{n_x}
\safemath{\zsig}{\bmz}
\safemath{\zsigdim}{\ysigdim}
\safemath{\rsig}{\bmr}
\safemath{\Adict}{\bA}
\safemath{\Adicttilde}{\widetilde{\Adict}}
\safemath{\Adictdim}{\outputdim\times\xsigdim}
\safemath{\avec}{\bma}
\safemath{\avectilde}{\tilde{\avec}}
\safemath{\Bdict}{\bB}
\safemath{\Bdicttilde}{\widetilde{\Bdict}}
\safemath{\Cdict}{\bC}
\safemath{\cvec}{\bmc}
\safemath{\Ddict}{\bD}
\safemath{\Ddictdim}{\ysigdim\times\xsigdim}
\safemath{\dvec}{\bmd}
\safemath{\Ddicttilde}{\widetilde{\bD}}
\safemath{\Bonb}{\bB}
\safemath{\bvec}{\bmb}
\safemath{\Bonbdim}{\ysigdim\times\ysigdim}
\safemath{\noise}{\bmn}
\safemath{\noisedim}{\ysigim}
\safemath{\err}{\bme}
\safemath{\errdim}{\ysigdim}
\safemath{\errset}{\setE}
\safemath{\nerr}{n_e}
\safemath{\delop}{\bP_\errset}
\safemath{\delopc}{\bP_{{\errset}^c}}

\safemath{\cplxi}{\imath}
\safemath{\cplxj}{\jmath}

\safemath{\dict}{\matD}
\safemath{\inputdim}{N}		
\safemath{\outputdim}{M}		
\safemath{\sparsity}{S}	
\safemath{\inputdimA}{{N_a}}	
\safemath{\inputdimB}{{N_b}}	
\safemath{\elemA}{{n_a}}	
\safemath{\elemB}{{n_b}}	
\safemath{\resA}{\matR_a}	
\safemath{\resB}{\matR_b}	
\safemath{\subD}{\matS} 
\safemath{\subA}{\matS_a} 
\safemath{\subB}{\matS_b} 
\safemath{\dicta}{\matA} 	
\safemath{\dictb}{\matB} 	
\safemath{\hollowS}{H}
\safemath{\hollowA}{H_a}
\safemath{\hollowB}{H_b}
\safemath{\cross}{Z}
\safemath{\coh}{\mu_d}			
\safemath{\coha}{\mu_a}			
\safemath{\cohb}{\mu_b}			
\safemath{\mubs}{\nu}	
\safemath{\cohm}{\mu_m} 
\safemath{\dictset}{\setD}	
\safemath{\dictsetp}{\dictset(\coh,\coha,\cohb)}	
\safemath{\dictsetgen}{\dictset_\text{gen}}
\safemath{\dictsetgenp}{\dictsetgen(\coh)}
\safemath{\dictsetonb}{\dictset_\text{onb}}
\safemath{\dictsetonbp}{\dictsetonb(\coh)}

\safemath{\leftside}{U}
\safemath{\rightsideA}{R_a}
\safemath{\rightsideB}{R_b}

\safemath{\indexS}{\setI_S} 

\safemath{\na}{n_a}			
\safemath{\nb}{n_b}			
\safemath{\coeffa}{p_i}	
\safemath{\coeffb}{q_j}	
\safemath{\seta}{\setP}		
\safemath{\setb}{\setQ}     
\safemath{\setw}{\setW}	
\safemath{\setz}{\setZ}	
\safemath{\cola}{\veca}		
\safemath{\colb}{\vecb}		
\safemath{\cold}{\vecd}		
\safemath{\inputvec}{\vecx} 	
\safemath{\error}{\vece}	
\safemath{\noiseout}{\vecz} 	
\safemath{\inputvecel}{x}
\safemath{\inputveca}{\vecx_a}
\safemath{\inputvecb}{\vecx_b}
\safemath{\outputvec}{\vecy}	
\safemath{\lambdamin}{\lambda_{\mathrm{min}}}

\safemath{\elltwo}{\ell_2}
\safemath{\ellone}{\ell_1}
\safemath{\ellzero}{\ell_0}
\safemath{\ellinf}{\ell_\infty}
\safemath{\ellinftilde}{\ell_{\widetilde\infty}}
\safemath{\licard}{Z(\coh,\coha,\cohb)}
\safemath{\xsol}{\hat{x}}
\safemath{\xbord}{x_b}		
\safemath{\xstat}{x_s}		
\safemath{\xstatLone}{\tilde{x}_s}
\safemath{\order}{\mathcal{O}} 
\safemath{\scales}{\Theta} 
\safemath{\ones}{\mathbf{1}} 
\safemath{\zeroes}{\mathbf{0}} 
\safemath{\thlone}{\kappa(\coh,\cohb)} 
\safemath{\constoneA}{\delta} 
\safemath{\constoneB}{\epsilon} 
\safemath{\nlarge}{L}				   
\safemath{\sumlarge}{S_\nlarge}
\safemath{\maxlarger}{P_\nlarge}	   
\safemath{\Pzero}{\textrm{P0}}	
\safemath{\Pone}{\textrm{P1}}
\safemath{\vecfir}{\vecw}			 
\safemath{\vecsec}{\vecz}
\safemath{\elvecfir}{w}              
\safemath{\elvecsec}{z}				 
\safemath{\nlargefir}{n}
\safemath{\normout}{\gamma}
\safemath{\auxfun}{h}
\safemath{\supp}{\textrm{supp}}

\safemath{\indexa}{\ell}
\safemath{\indexb}{r}
\safemath{\indexc}{i}
\safemath{\indexd}{j}

\safemath{\project}{P}

%% file: 15TCOM_qmmimo_final.bbl
\begin{thebibliography}{10}
\providecommand{\url}[1]{#1}
\csname url@samestyle\endcsname
\providecommand{\newblock}{\relax}
\providecommand{\bibinfo}[2]{#2}
\providecommand{\BIBentrySTDinterwordspacing}{\spaceskip=0pt\relax}
\providecommand{\BIBentryALTinterwordstretchfactor}{4}
\providecommand{\BIBentryALTinterwordspacing}{\spaceskip=\fontdimen2\font plus
\BIBentryALTinterwordstretchfactor\fontdimen3\font minus
  \fontdimen4\font\relax}
\providecommand{\BIBforeignlanguage}[2]{{%
\expandafter\ifx\csname l@#1\endcsname\relax
\typeout{** WARNING: IEEEtran.bst: No hyphenation pattern has been}%
\typeout{** loaded for the language `#1'. Using the pattern for}%
\typeout{** the default language instead.}%
\else
\language=\csname l@#1\endcsname
\fi
#2}}
\providecommand{\BIBdecl}{\relax}
\BIBdecl

\bibitem{murmann14-11a}
\BIBentryALTinterwordspacing
B.~Murmann, ``{ADC Performance Survey 1997-2014}.'' [Online]. Available:
  \url{http://web.stanford.edu/~murmann/adcsurvey.html}
\BIBentrySTDinterwordspacing

\bibitem{Marzetta10}
T.~L. Marzetta, ``Non-cooperative cellular wireless with unlimited numbers of
  base station antennas,'' \emph{IEEE Trans. Wireless Comm.}, vol.~9, no.~11,
  pp. 3590--3600, Nov. 2010.

\bibitem{larsson14-02a}
E.~G. Larsson, F.~Tufvesson, O.~Edfors, and T.~L. Marzetta, ``Massive {MIMO}
  for next generation wireless systems,'' \emph{{IEEE} Commun. Mag.}, pp.
  186--195, Feb. 2014.

\bibitem{bjornson14-11a}
E.~Bj{\"o}rnson, J.~Hoydis, M.~Kountouris, and M.~Debbah, ``Massive {MIMO}
  systems with non-ideal hardware: Energy efficiency, estimation, and capacity
  limits,'' \emph{{IEEE} Trans. Inf. Theory}, vol.~11, no.~60, pp. 7112--7139,
  Nov. 2014.

\bibitem{Studer_Tx_OFDM}
C.~Studer, M.~Wenk, and A.~Burg, ``{MIMO} transmission with residual
  transmit-{RF} impairments,'' in \emph{Int. ITG Workshop on Smart Antennas
  (WSA)}, Bremen, Germany, Feb. 2010, pp. 189--196.

\bibitem{gustavsson14-12a}
U.~Gustavsson, C.~Sanch\'ez-Perez, T.~Eriksson, F.~Athley, G.~Durisi,
  P.~Landin, K.~Hausmair, C.~Fager, and L.~Svensson, ``On the impact of
  hardware impairments on massive {MIMO},'' in \emph{Proc. IEEE Global
  Telecommun. Conf. (GLOBECOM)}, Austin, TX, Dec. 2014, pp. 294--300.

\bibitem{RPL14}
\BIBentryALTinterwordspacing
C.~Risi, D.~Persson, and E.~G. Larsson, ``Massive {MIMO} with 1-bit {ADC},''
  Apr. 2014. [Online]. Available: \url{http://arxiv.org/abs/1404.7736}
\BIBentrySTDinterwordspacing

\bibitem{jacobsson15-06a}
S.~Jacobsson, G.~Durisi, M.~Coldrey, U.~Gustavsson, and C.~Studer, ``One-bit
  massive {MIMO}: channel estimation and high-order modulations,'' in
  \emph{Proc. IEEE Int. Conf. Commun. (ICC)}, London, U.K., Jun. 2015.

\bibitem{davenport14-a}
M.~A. Davenport, Y.~Plan, E.~{van den Berg}, and M.~Wootters, ``1-{Bit} matrix
  completion,'' \emph{Information and Inference}, vol.~3, no. 189-223, July
  2014.

\bibitem{knudson14-04a}
\BIBentryALTinterwordspacing
K.~Knudson, R.~Saab, and R.~Ward, ``One-bit compressive sensing with norm
  estimation,'' Apr. 2014. [Online]. Available:
  \url{http://arxiv.org/abs/1404.6853}
\BIBentrySTDinterwordspacing

\bibitem{liang15-04a}
N.~Liang and W.~Zhang, ``Mixed-{ADC} massive {MIMO},'' \emph{{IEEE} J. Sel.
  Areas Commun.}, 2016, to appear.

\bibitem{singh09-12a}
J.~Singh, O.~Dabeer, and U.~Madhow, ``On the limits of communication with
  low-precision analog-to-digital conversion at the receiver,'' \emph{{IEEE}
  Trans. Commun.}, vol.~57, no.~12, pp. 3629--3639, Dec. 2009.

\bibitem{mezghani07-06a}
A.~Mezghani and J.~Nossek, ``On ultra-wideband {MIMO} systems with 1-bit
  quantized outputs: Performance analysis and input optimization,'' in
  \emph{Proc. IEEE Int. Symp. Inf. Theory (ISIT)}, Nice, France, Jun. 2007, pp.
  1286--1289.

\bibitem{mezghani08-07a}
------, ``Analysis of {Rayleigh}-fading channels with 1-bit quantized output,''
  in \emph{Proc. IEEE Int. Symp. Inf. Theory (ISIT)}, Toronto, ON, Canada, Jul.
  2008, pp. 260--264.

\bibitem{mezghani09-06a}
------, ``Analysis of 1-bit output noncoherent fading channels in the low {SNR}
  regime,'' in \emph{Proc. IEEE Int. Symp. Inf. Theory (ISIT)}, Seoul, Korea,
  Jun. 2009, pp. 1080--1084.

\bibitem{mo14-10a}
J.~Mo and R.~W. {Heath Jr}, ``Capacity analysis of one-bit quantized {MIMO}
  systems with transmitter channel state information,'' \emph{{IEEE} Trans.
  Signal Process.}, vol.~63, no.~20, pp. 5498 -- 5512, Oct. 2015.

\bibitem{koch13-09a}
T.~Koch and A.~Lapidoth, ``At low {SNR}, asymmetric quantizers are better,''
  \emph{{IEEE} Trans. Inf. Theory}, vol.~59, no.~9, pp. 5421--5445, Sep. 2013.

\bibitem{verdu02-06a}
S.~Verd{\'u}, ``Spectral efficiency in the wideband regime,'' \emph{{IEEE}
  Trans. Inf. Theory}, vol.~48, no.~6, pp. 1319--1343, Jun. 2002.

\bibitem{narasimha12-07a}
R.~Narasimha, M.~Lu, N.~Shanbhag, and A.~Singer, ``{BER}-optimal
  analog-to-digital converters for communication links,'' \emph{{IEEE} Trans.
  Signal Process.}, vol.~60, no.~7, pp. 3683--3691, Jul. 2012.

\bibitem{zeitler12-09a}
G.~Zeitler, A.~Singer, and G.~Kramer, ``Low-precision {A/D} conversion for
  maximum information rate in channels with memory,'' \emph{{IEEE} Trans.
  Commun.}, vol.~60, no.~9, pp. 2511--2521, Sep. 2012.

\bibitem{shamai-shitz94-06a}
S.~Shamai~(Shitz), ``Information rates by oversampling the sign of a
  bandlimited process,'' \emph{{IEEE} Trans. Inf. Theory}, vol.~40, no.~4, pp.
  1230--1236, Jun. 1994.

\bibitem{koch10-11a}
T.~Koch and A.~Lapidoth, ``Increased capacity per unit-cost by oversampling,''
  in \emph{IEEE 26th Convention of Electrical and Electronics Engineers in
  Israel (IEEEI),}, Eliat, Israel, Nov. 2010, pp. 684 --688.

\bibitem{zhang12-02a}
W.~Zhang, ``A general framework for transmission with transceiver distortion
  and some applications,'' \emph{{IEEE} Trans. Commun.}, vol.~60, no.~2, pp.
  384--399, Feb. 2012.

\bibitem{krone12-05a}
S.~Krone and G.~Fettweis, ``Capacity of communications channels with 1-bit
  quantization and oversampling at the receiver,'' in \emph{{IEEE} Sarnoff
  Symp. (SARNOFF)}, Newark, NJ, May 2012.

\bibitem{dabeer06-12a}
O.~Dabeer and A.~Karnik, ``Signal parameter estimation using 1-bit dithered
  quantization,'' \emph{{IEEE} Trans. Inf. Theory}, vol.~52, no.~12, pp.
  5389--5405, Dec. 2006.

\bibitem{mezghani10-02a}
A.~Mezghani, F.~Antreich, and J.~A. Nossek, ``Multiple parameter estimation
  with quantized channel output,'' in \emph{ITG Smart Antenna Workshop},
  Bremen, Germany, Feb. 2010, pp. 143--150.

\bibitem{zeitler12-06a}
G.~Zeitler, G.~Kramer, and A.~Singer, ``Bayesian parameter estimation using
  single-bit dithered quantization,'' \emph{{IEEE} Trans. Signal Process.},
  vol.~60, no.~6, pp. 2713--2726, Jun. 2012.

\bibitem{zymnis2010}
A.~Zymnis, S.~Boyd, and E.~Cand\'es, ``Compressed sensing with quantized
  measurements,'' \emph{IEEE Signal Process. Lett.}, vol.~17, no.~2, pp.
  149--152, Feb. 2010.

\bibitem{boufounos08-03a}
P.~T. Boufounos and R.~G. Baraniuk, ``1-{Bit} compressive sensing,'' in
  \emph{Proc. Annual Conf. Inf. Sciences Syst. (CISS)}, Mar. 2008, pp. 16--21.

\bibitem{plan13-a}
Y.~Plan and R.~Vershynin, ``One-bit compressed sensing by linear programming,''
  \emph{Comm. Pure Appl. Math.}, vol.~66, no.~8, pp. 1275--1297, Feb. 2013.

\bibitem{lok98-08a}
T.~Lok and V.-W. Wei, ``Channel estimation with quantized observation,'' in
  \emph{Proc. IEEE Int. Symp. Inf. Theory (ISIT)}, Cambridge, MA, Aug. 1998, p.
  333.

\bibitem{ivrlac07-02a}
M.~T. Ivrlac and J.~A. Nossek, ``On {MIMO} channel estimation with single-bit
  signal quantization,'' in \emph{ITG Smart Antenna Workshop}, Vienna, Austria,
  Feb. 2007.

\bibitem{wang14-04a}
S.~Wang, Y.~Li, and J.~Wang, ``Multiuser detection in massive spatial
  modulation {MIMO} with low-resolution {ADCs},'' \emph{{IEEE} Trans. Wireless
  Commun.}, vol.~14, no.~4, pp. 2156--2168, Apr. 2014.

\bibitem{mezghani2010belief}
A.~Mezghani and J.~A. Nossek, ``Belief propagation based {MIMO} detection
  operating on quantized channel output,'' in \emph{Proc. IEEE Int. Symp. Inf.
  Theory (ISIT)}, Austin, TX, Jun. 2010, pp. 2113--2117.

\bibitem{mezghani12-03a}
A.~Mezghani, M.~Rouatbi, and J.~Nossek, ``An iterative receiver for quantized
  {MIMO} systems,'' in \emph{{IEEE} Mediterranean Electrotechnical Conf.
  {(MELECOM)}}, Hammamet, Tunisia, Mar. 2012, pp. 1049--1052.

\bibitem{choi15-07a}
J.~Choi, J.~Mo, and R.~W. {Heath Jr}, ``Near maximum-likelihood detector and
  channel estimator for uplink multiuser massive {MIMO} systems with one-bit
  {ADCs},'' \emph{{IEEE} Trans. Commun.}, 2016, to appear.

\bibitem{Bellasi13}
D.~Bellasi, L.~Bettini, C.~Benkeser, T.~Burger, Q.~Huang, and C.~Studer,
  ``{VLSI} design of a monolithic compressive-sensing wideband
  analog-to-information converter,'' \emph{IEEE J. Emerg. Sel. Topics Circuits
  Syst.}, vol.~3, no.~4, pp. 552--565, Mar. 2013.

\bibitem{ganti00-11a}
A.~Ganti, A.~Lapidoth, and I.~Telatar, ``Mismatched decoding revisited: general
  alphabets, channels with memory, and the wide-band limit,'' \emph{{IEEE}
  Trans. Inf. Theory}, vol.~46, no.~7, pp. 2315--2328, Nov. 2000.

\bibitem{SJSB11}
D.~Seethaler, J.~Jald{\'e}n, C.~Studer, and H.~B\"olcskei, ``On the complexity
  distribution of sphere decoding,'' \emph{{IEEE} Trans. Inf. Theory}, vol.~57,
  no.~9, pp. 5754--5768, Sept. 2011.

\bibitem{JO05}
J.~Jalden and B.~Ottersten, ``On the complexity of sphere decoding in digital
  communications,'' \emph{{IEEE} Trans. Signal Process.}, vol.~53, no.~4, pp.
  1474--1484, Apr. 2005.

\bibitem{studer2011asic}
C.~Studer, S.~Fateh, and D.~Seethaler, ``{ASIC} implementation of soft-input
  soft-output {MIMO} detection using {MMSE} parallel interference
  cancellation,'' \emph{{IEEE} J. Solid-State Circuits}, vol.~46, no.~7, pp.
  1754--1765, Jul. 2011.

\bibitem{seethaler2004efficient}
D.~Seethaler, G.~Matz, and F.~Hlawatsch, ``An efficient {MMSE}-based
  demodulator for {MIMO} bit-interleaved coded modulation,'' in \emph{Proc.
  IEEE Global Telecommun. Conf. (GLOBECOM)}, vol.~4, Dallas, TX, Nov. 2004, pp.
  2455--2459.

\bibitem{WYWDCS2014}
M.~Wu, B.~Yin, G.~Wang, C.~Dick, J.~Cavallaro, and C.~Studer, ``Large-scale
  {MIMO} detection for {3GPP LTE}: Algorithm and {FPGA} implementation,''
  \emph{{IEEE} J. Sel. Topics Signal Process.}, vol.~8, no.~5, pp. 916--929,
  Oct. 2014.

\bibitem{yin2014conjugate}
B.~Yin, M.~Wu, J.~R. Cavallaro, and C.~Studer, ``Conjugate gradient-based
  soft-output detection and precoding in massive {MIMO} systems,'' in
  \emph{Proc. IEEE Global Telecommun. Conf. (GLOBECOM)}, Austin, TX, Dec. 2014,
  pp. 3696--3701.

\bibitem{IEEE11n}
\emph{{IEEE} Draft Standard; Part 11: Wireless LAN Medium Access Control (MAC)
  and Physical Layer (PHY) specifications; Amendment 4: Enhancements for Higher
  Throughput}, P802.11n/D3.0, Sep. 2007.

\bibitem{schenk2005performance}
T.~C. Schenk, P.~F. Smulders, and E.~R. Fledderus, ``Performance of {MIMO OFDM}
  systems in fading channels with additive {TX} and {RX} impairments,'' in
  \emph{Proc. IEEE BENELUX/DSP Valley Signal Process. Symp.}, Antwerp, Belgium,
  Apr. 2005, pp. 41--44.

\bibitem{schenk2008rf}
T.~C. Schenk, \emph{{RF} imperfections in high-rate wireless systems: impact
  and digital compensation}.\hskip 1em plus 0.5em minus 0.4em\relax Dordrecht,
  The Netherlands: Springer, 2008.

\bibitem{SL12}
C.~Studer and E.~G. Larsson, ``{PAR}-aware large-scale multi-user {MIMO-OFDM}
  downlink,'' \emph{IEEE J. Sel. Areas Comm.}, vol.~31, no.~2, pp. 303--313,
  Feb. 2013.

\bibitem{bolcskei2002capacity}
H.~B\"olcskei, D.~Gesbert, and A.~Paulraj, ``On the capacity of {OFDM}-based
  spatial multiplexing systems,'' \emph{{IEEE} Trans. Commun.}, vol.~50, no.~2,
  pp. 225--234, Feb. 2002.

\bibitem{haene2008ofdm}
S.~Haene, A.~Burg, N.~Felber, and W.~Fichtner, ``{OFDM} channel estimation
  algorithm and {ASIC} implementation,'' in \emph{4th European Conference on
  Circuits and Systems for Communications (ECCSC 2008)}, 2008, pp. 270--275.

\bibitem{paulraj03}
A.~Paulraj, R.~Nabar, and D.~Gore, \emph{Introduction to Space-Time Wireless
  Communications}.\hskip 1em plus 0.5em minus 0.4em\relax Cambridge Univ.
  Press, 2003.

\bibitem{GSB14}
\BIBentryALTinterwordspacing
T.~Goldstein, C.~Studer, and R.~G. Baraniuk, ``A field guide to
  forward-backward splitting with a {FASTA} implementation,'' Nov. 2014.
  [Online]. Available: \url{http://arxiv.org/abs/1411.3406}
\BIBentrySTDinterwordspacing

\bibitem{BT09}
A.~Beck and M.~Teboulle, ``A fast iterative shrinkage-thresholding algorithm
  for linear inverse problems,'' \emph{SIAM J. Imag. Sci.}, vol.~2, no.~1, pp.
  183--202, Jan. 2009.

\bibitem{max60-03a}
J.~Max, ``Quantizing for optimal distortion,'' \emph{{IRE} Trans. Info.
  Theory}, vol.~6, no.~1, pp. 7--12, Mar. 1960.

\bibitem{lloyd82-03a}
S.~P. Lloyd, ``Least squares quantization in pcm,'' \emph{{IEEE} Trans. Inf.
  Theory}, vol.~28, no.~2, pp. 129--137, Mar. 1982.

\bibitem{GSB15software}
\BIBentryALTinterwordspacing
T.~Goldstein, C.~Studer, and R.~G. Baraniuk, ``{FASTA:} {A} generalized
  implementation of forward-backward splitting,'' Aug. 2015. [Online].
  Available: \url{http://arxiv.org/abs/1501.04979}
\BIBentrySTDinterwordspacing

\bibitem{SB10}
C.~Studer and H.~B\"olcskei, ``Soft--input soft--output single tree-search
  sphere decoding,'' \emph{{IEEE} Trans. Inf. Theory}, vol.~56, no.~10, pp.
  4827--4842, Oct. 2010.

\bibitem{studer2008soft}
C.~Studer, A.~Burg, and H.~B\"olcskei, ``Soft-output sphere decoding:
  Algorithms and {VLSI} implementation,'' \emph{{IEEE} J. Sel. Areas Commun.},
  vol.~26, no.~2, pp. 290--300, Feb. 2008.

\bibitem{3GPPLTE}
\emph{3rd Generation Partnership Project; Technical Specification Group Radio
  Access Network; Evolved Universal Terrestrial Radio Access (E-UTRA);
  Multiplexing and channel coding (Release 9)}, 3GPP Organizational Partners TS
  36.212, Rev. 8.3.0, May 2008.

\end{thebibliography}
